\documentclass{aa}  

\usepackage[colorlinks]{hyperref}
\usepackage{natbib}
\usepackage{scrhack} 
\hypersetup{
    colorlinks=true,
    linkcolor=[RGB]{0,0,255}, 
    linkbordercolor={0 0 1},
    filecolor=[RGB]{0,0,255},      
    urlcolor=[RGB]{0,0,255},
    citecolor=[RGB]{0,0,255},
}
\usepackage{amsmath}
\usepackage{amssymb}
\usepackage{physics}
\usepackage{xcolor} 
\newcommand{\degree}{^\circ}

\usepackage{graphicx}
\usepackage{float}
\usepackage{adjustbox}

\newcommand{\fix}{}

\usepackage{txfonts}

\begin{document}

    \title{\texttt{ROLLIN’}: Rotating Globular Cluster Simulations} 
    \subtitle{II. The complex morphology of globular clusters driven by multi-scale dynamics}

   \author{Arn Marklund
          \inst{1}\thanks{   \email{arn.marklund@astro.unistra.fr}}
          \and Paolo Bianchini \inst{1}
          \and Anna Lisa Varri \inst{2}\fnmsep \inst{3}
          \and Katarina Kraljic \inst{1}
          \and Giulia Pagnini \inst{1}
          }
          
   \institute{Université de Strasbourg, CNRS, Observatoire astronomique de Strasbourg, UMR 7550, F-67000 Strasbourg, France 
         \and School of Mathematics and Maxwell Institute for Mathematical Sciences, University of Edinburgh, Kings Buildings, Edinburgh EH9 3FD, UK 
         \and Institute for Astronomy, University of Edinburgh, Royal Observatory, Blackford Hill, Edinburgh EH9 3HJ, UK 
             }

   \date{Received May 15, 2026 ; accepted July 30, 2026}

  \abstract
   {Globular clusters (GCs) are inherently non-spherical systems that in many cases show internal rotation. Typically, rotation is considered as the main driver of a GC's morphology; however, the relationship between ellipticity and rotational support is not a simple one-to-one mapping, and other multi-scale dynamical processes could contribute.}
   {We aim to provide a comprehensive interpretation of how morphology evolves in realistic models of rotating GCs, and how it correlates with key physical ingredients and processes, including mass loss, stellar evolution, external tidal field, and two-body relaxation.}
   {Using the \texttt{ROLLIN’} suite of direct N-body simulations, we compare two classical methods for measuring a GC's shape: iso-density contours and the second-moment tensor method, and quantify their differences in measured ellipticity. We then measure the intrinsic ellipticity and triaxiality of our models using the tensor method, and explore how these quantities change with time and the main physical mechanisms that drive them.}
   {We find that the early evolution of GCs can be dominated by dynamical instabilities driven by internal rotation and velocity anisotropy, leading to the formation of bar-like structures that rapidly erode due to collisional effects around the time of the first core collapse. These bars are stronger and more long-lived ($\lesssim 800$ Myr) in strongly rotating clusters with longer relaxation times and subject to stellar evolution. In the long term, clusters evolve toward a less flattened and gradually triaxial configuration. This trend is more pronounced for clusters that experience stronger mass loss, that are more tidally filling and isotropic, and with lower rotational support. The inner ($<r_{50\%}$) and outer ($>r_{50\%}$) regions also show different morphologies; the inner parts of our models remain oblate and are dominated by internal rotation, while the outer region is triaxial and resembles the shape of the Roche lobe due to the external tidal field.
   }
   {Our models provide a physical explanation for the observational \(V/\sigma\)–ellipticity relation and demonstrate that morphology can serve as a \fix{reliable tracer of the dynamical state of the GC}. Initially retrograde, dense, and inclined rotating models deviate systematically from the overall trend, providing physically motivated explanations for observational outliers. Upcoming large-scale photometric surveys will significantly benefit from comprehensive physical frameworks capable of interpreting GC evolution in terms of morphology and kinematics. 
   }

   \keywords{methods: numerical -- globular clusters: general, galaxies: star clusters: general}

   \maketitle

\section{Introduction}

\begin{table*}[htbp!]
\centering
\caption{Initial conditions for simulations not included in Paper I.}
\begin{adjustbox}{max width=\textwidth}
\label{tab: ICs}
\begin{tabular}{lcccccccccl}  
\hline \hline
Name & $N$ & $M$ & $r_{50\%}$ & $\Sigma_*$ & $d$ & $r_{50\%}/r_j$ & $V_{\rm peak}$ & $V_{\rm peak}/\sigma_0$ & $\log(t_{\rm rh}/\mathrm{yr})$ & comments \\
 & $10^5$ & $10^5\,\mathrm{M}_\odot$ & pc & $10^3\,\mathrm{M}_\odot\,\mathrm{pc}^{-2}$ & kpc &  & km/s &  &  & \\
\hline
 \texttt{250k-A-R4-25-incl45} & 2.5 & $1.44$ & 4 & 1.5 & 25 & 0.02 & 5.64 & 1.22  & 8.97 &  rot-angle 45 deg\\
 \texttt{250k-W6-R4-25-norot} &  2.5 & 1.43 & 4 & 1.4 & 25 & 0.02 & -- & -- & 8.98 & no rotation\\
\hline
\end{tabular}
\end{adjustbox}
\tablefoot{The initial conditions of these simulations follow the definitions introduced in Paper~I. See Paper~I for a complete description of the initial conditions and the listed quantities.}
\end{table*}

\begin{table*}[htbp!]
\centering
\caption{Properties at 12 Gyr for the \texttt{ROLLIN'} models in Table \ref{tab: ICs}.}
\begin{adjustbox}{max width=\textwidth}
\label{tab: 12Gyr}
\begin{tabular}{lcccccccccl} 
\hline \hline
Name & $N$ & $M$ & $\Delta M/M$ & $r_{50\%}$ & $\Sigma_*$ & $r_{50\%}/r_j$ & $V_{\rm peak}$ & $V_{\rm peak}/\sigma_0$ & $\beta_{<50\%}$ & $\log(t_{\rm rh}/\mathrm{yr})$ \\
 & $10^5$ & $10^5\,\mathrm{M}_\odot$ & & pc & $10^3\,\mathrm{M}_\odot\,\mathrm{pc}^{-2}$ & & km/s &  &  & \\
\hline
 \texttt{250k-A-R4-25-incl45} &  2.35 & 0.79 & 0.45 & 13.7 & 0.07 & 0.1 & 1.15 & 0.47 & 0.18 & 9.96  \\
 \texttt{250k-W6-R4-25-norot} &  2.32 & 0.78 & 0.46 & 16.3 & 0.05 & 0.12 & -0.17 & 0.07 & 0.17 & 10.07 \\
\hline
\end{tabular}
\end{adjustbox}
\tablefoot{The definitions of the listed quantities are given in Paper~I.}
\end{table*}

Observations over the past two decades have revealed that many Galactic globular clusters (GCs) show measurable internal rotation \citep[e.g.][see also \citealp{Bianchini+2026} for a recent overview]{Bellazzini+2012,Bianchini+13,Bianchini+2018,Fabricius+2014,Lardo+2015,Kamann+2018,VasilievBaumgardt2021,Petralia+2024,Leitinger+2025}. This finding suggests that rotation may be a nearly universal property of GCs, at least at some stage of their evolution. However, direct kinematic measurements are challenging for distant clusters, limiting the study of detailed rotational profiles mostly to Milky Way GCs. 

It has also since long been established that GCs are not strictly spherical, but show measurable deviations from an axial ratio of unity \citep{PeaseShapley1917}. Many early studies of star clusters in the Milky Way, the Magellanic Clouds, and selected local dwarf galaxies have suggested that a main driver of this deviation, or flattening, is rotation \citep[see e.g.][]{FrenkFall1982,Geyer+1983,Webbink1985,WhiteShawl1987,Kontizas+1989,Kontizas+1990,LagouteLongaretti1996,LongarettiLagoute1996,ChenChen2010,Lahen+2020,ReyesAnderson2024,Freour+2026}, which would suggest that the shape of a GC provides a useful, though indirect, approach for studying their kinematics \citep[see also][]{DavoustPrugniel1990}. \citet{Freour+2026} recently derived robust ellipticity measurements for 29 Galactic GCs, finding that the flattening for a majority of GCs can be attributed to their internal rotation. However, they also found a few non-rotating GCs to have a significant ellipticity, suggesting that the flattening of GCs is the result of multi-physical processes \citep[see also][]{vandenBergh2008,ChenChen2010}.

Indeed, structural flattening may also be attributed to the gravitational influence of the host galaxy through tidal forces \citep{BertinVarri2008,VarriBertin2009,Carretta+2010,Kupper+2010,ArditiVarri2026}, as well as to anisotropy in the velocity space \citep{WhiteShawl1987,Fiestas+2006}. The combination of internal kinematic properties and the external environment are likely responsible for measured radial variations in a GC's ellipticity \citep{Geyer+1983}, where the more central and outer regions may appear more circular \citep[see for example M15,][]{petkova+2026}. It is, as such, possible that different processes simultaneously influence different regions of the GC, resulting in a non-trivial morphology deviating from simple axisymmetric configurations. Distinguishing between all these drivers, and tracing how they evolve in tandem, is key to properly interpreting observed GC morphologies. 

Previous theoretical studies have addressed the interplay between internal rotation, external tides, and morphological evolution using $N$-body simulations \citep[e.g.][see also \citealp{Bissekenov+2025,Kamlah+2022}]{Tiongco+2016b,Tiongco+2016a,Tiongco+18,Tiongco+2022}, or equilibrium models \citep[e.g.][]{VarriBertin2012}, or Fokker-Planck approaches \citep[e.g.][]{EinselSpurzem1999}. These investigations show that GCs gradually lose angular momentum through a combination of two-body relaxation and mass loss due to evolution within a tidal field, resulting in a steady decline for both rotational velocity and flattening. Structurally, these dynamical processes produce complex morphological signatures. Idealised simulations by \citet{Tiongco+2022} demonstrated that rotating clusters develop triaxial shapes where ellipticity and axis orientation vary with radius, particularly for systems with strong initial rotation. Interestingly, when a cluster’s rotation axis initially is misaligned with its orbital angular momentum, tidal torques can induce a precession and nutation of the internal rotation axis \citep{Tiongco+18,Tiongco+2022}, further enriching the spectrum of complex morphologies that a cluster can display. Furthermore, the authors found that the orientation of the minor axis does not always coincide with the rotation axis; a misalignment seen in nuclear star clusters and elliptical galaxies \citep[e.g.][]{Seth+2010,Emsellem+2011}. The presence of such misalignment in GCs would point to a more intricate coupling between rotation, tidal interactions, and anisotropy.

Stellar evolution adds yet another component to a GC's evolution. \citet{Kamlah+2022} examined rotating GC models both with and without stellar evolution, and found that mass segregation and core collapse drive transient shifts in morphology. More specifically, during the early core collapse, they found that clusters tend to become more spherical and isotropic, while the formation of bars via dynamical instabilities briefly increased triaxiality in the very early phases, but only when stellar evolution was enabled. After these phases, the systems evolve toward a restored axisymmetric equilibrium.

With upcoming large scale surveys, such as Euclid, the Rubin Observatory (LSST) and the Roman Space Telescope, understanding and quantifying the morphologies of GCs is becoming more important \citep{Massari+2025}, in particular where kinematical data is sparse. For example, current estimates suggest that galaxies within 100 Mpc in the Euclid footprint may host on the order of \(8\times10^{5}\) GCs, of which \(\sim3.5\times10^{5}\) will be within the survey’s detection limits \citep{Voggel+2025}. These facilities will enable statistical studies of GCs and their morphologies far beyond the Local Group. As such, they are also underlining the need for robust physical frameworks capable of interpreting observed morphologies with respect to rotation, tidal fields, anisotropy, and a cluster's dynamical evolution. The development of such a framework is, therefore, timely and essential for exploiting the full scientific potential of forthcoming next-generation surveys. Approaches combining $N$-body simulations and deep learning are promising tools for a scalable solution to large datasets (Marklund et al., in prep.).

In the first paper of this series \citep[][hereafter Paper I]{Bianchini+2026}, we presented the \texttt{ROLLIN’} (ROtating globular cLusters Long-term evolutioN) simulation suite—25 direct N-body models with stellar counts between $2.5\times10^5$ and $1.5\times10^6$, computed using \texttt{NBODY6++GPU} \citep{Aarseth2003,Wang+15}. \fix{Using such high stellar counts is essential to reproduce the collisional dynamics of globular clusters while consistently modelling stellar evolutionary effects.} That work focused on the long-term kinematic evolution and angular momentum transport within rotating clusters, showing that the decay of rotation and anisotropy correlates closely with the fraction of mass lost over time. 

This study constitutes part two of the series, and aims to explore an in-depth morphological characterisation of our suite of GC simulations. In particular, we explore the morphology of GCs both in the early time and in the long term with respect to the evolution of mass loss, the tidal field, anisotropy, and internal rotation, and we provide a comparison of our models to state-of-the-art measurements for Galactic GCs. The rest of the paper is structured as follows: in Sect. \ref{subsec: Nbody simulations} we give a short summary of the $N$-body models we consider, and in Sect. \ref{Ellipticity and Triaxiality} we explore different approaches for determining a GC's morphology; in particular, ellipticity and triaxiality. We look at how these quantities evolve with time, and how they correlate with other dynamical and kinematical quantities in Sect. \ref{sect: Time evolution}, while explicitly considering the impact of the tidal field in Sect. \ref{sect: Tidal field}. We compare the morphologies of our simulations to a subset of Galactic GCs in Sect. \ref{sect: Observations}. Finally, our conclusions are laid out in Sect. \ref{Conclusions}.

\section{Method}

\subsection{$N$-body simulations} \label{subsec: Nbody simulations}

Paper I introduced a suite of 25 realistic and axisymmetric $N$-body simulations with different initial rotation strengths \citep[denoted by A,B,C, and quantified by $\hat\omega = 0.3,0.2,0.1$, respectively, see also][]{VarriBertin2012}, number of particles (250k-1.5M), half-mass radii (1-4 pc), and distances to the centre of the external potential (5-25 kpc). These models are all initially axisymmetric. In this paper we extend this suite with two new simulations that have similar initial properties to those first reported in Paper I, which are summarised in Table \ref{tab: ICs}. We also include their corresponding properties at 12 Gyr in Table \ref{tab: 12Gyr}.

The first new simulation, \texttt{250k-A-R4-25-incl45}, extends our most common initial conditions (250k particles, a rotation strength of $\hat\omega = 0.3$, a half-mass radius of $4$ pc, and a circular orbit at $25$ kpc), but relaxes the assumption of a rotation axis aligned with the orbital angular momentum vector. We pick an initial inclination angle of $45\degree$, and it is therefore the first model in our suite that is not strictly prograde or retrograde. 

We also introduce a non-rotating spherical \citet{Wilson1975} model, \texttt{250k-W6-R4-25-norot}, as a control simulation. The suite contains two other spherical Wilson models (described in paper I) which have their rotation introduced via the "Lynden-Bell demon" \citep{Lynden-Bell1962}. One of the main motivations for including these non-standard models, is to explore other initial prescriptions for rotation that does not assume that GCs form axisymmetrically. Hereafter, we refer to these as the spherical Wilson models.
 
\subsection{Characterising ellipticity and triaxiality} \label{Ellipticity and Triaxiality}

The morphology of a stellar system, such as a GC, can be quantified in several ways depending on the available data and the physical scale of interest. Common approaches include fitting iso-density contours \citep[e.g.][]{Staneva+1996,Stetson+2019} or computing the second moment tensor of the stellar mass distribution \citep[e.g.][]{FallFrenk1983}. Although both techniques aim to measure the degree of flattening or ellipticity, they emphasise different aspects of the structure and convey slightly different physical interpretations. We explore the two methodologies in this section using:
\begin{enumerate}
	\item[(i)] ellipse fitting in projected density maps (iso-density contours)
	\item[(ii)] the second moment tensor on the 3D distribution of stars.
\end{enumerate}

In the first approach (hereafter iso-density method), the ellipticity is derived from the projected geometry of surfaces of constant surface density (or brightness). This method is directly comparable to observational data, as it traces the projected ellipticity of the cluster in the plane of the sky. It reflects how the surface brightness, or surface stellar density, is distributed as a function of radius, allowing an assessment of radial variations in cluster shape. However, such measurements are inherently two-dimensional projections of a three-dimensional system. The loss of information along the line of sight means that intrinsic structures (such as internal anisotropies or variations in the vertical density profile) are not fully captured. This projection effect can bias the observed ellipticity, particularly in dense central regions where overlapping stellar populations may obscure true spatial variations. 

We use surface density maps in this approach and compute 2D histograms of the simulations projected along the main cartesian coordinate axes. The axes are defined by $\hat x$ pointing away from the external potential, $\hat z$ pointing along the orbital angular momentum vector\footnote{For all models, except \texttt{250k-A-R4-25-incl45}, the $z$-axis coincides with the rotation axis.}, and $\hat y$ pointing in the direction of the orbit of the GC (perpendicular to both $\hat x$ and $\hat z$). The major-, and minor axes $a$ and $b$, as well as the position angle $\theta$ can be determined by numerically solving for

\begin{equation} \label{eqn: rotated ellipse}
f(x, y) = \frac{(x \cos\theta + y \sin\theta)^2}{a^2} + \frac{(-x \sin\theta + y \cos\theta)^2}{b^2} - 1
\end{equation}
which represents the equation of a rotated ellipse, along two coordinates $x,~y$. The ellipticity is then defined as
\begin{equation} \label{eqn: ellipticity}
e = 1 - \frac{b}{a}.
\end{equation}
We consider the projections in the $xy-$, $xz-$, and $yz-$, planes, which provide three different values of ellipticity, but other arbitrary projections are also possible.

\begin{figure*}[htbp!]
    \centering
    \includegraphics[width=\linewidth]{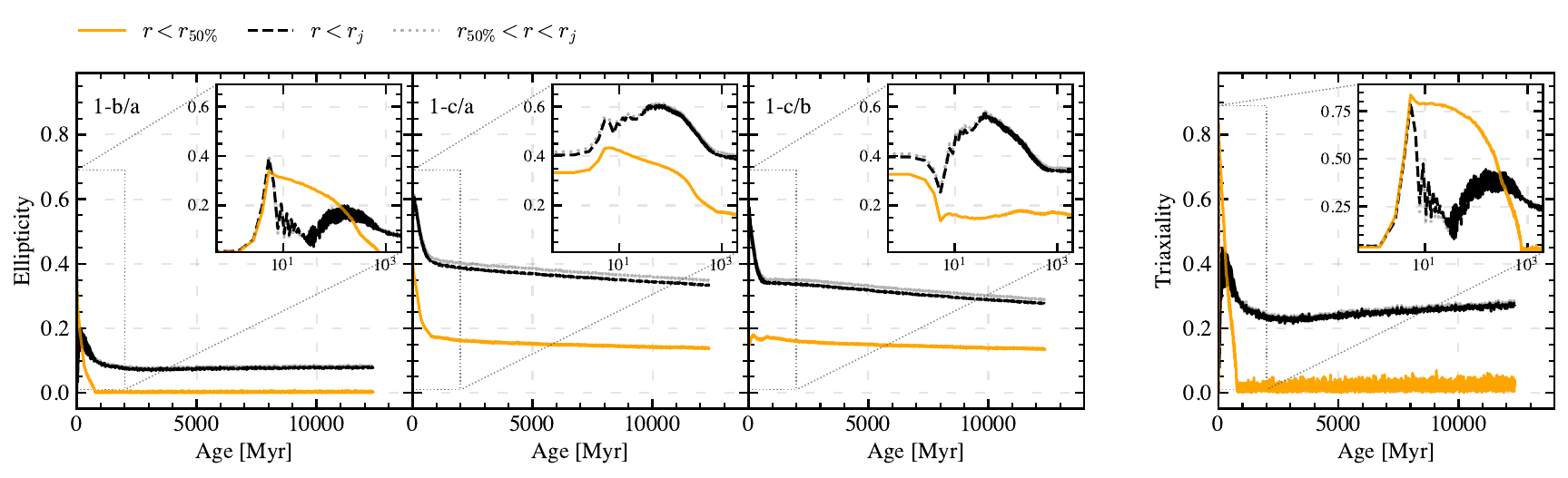} 
    \caption{Time evolution of \texttt{1.5M-A-R4-10} and its ellipticity for the ratios $b/a$, $c/a$, and $c/b$, and its triaxiality $T$, as shown in the four panels, respectively. The inset axes show the evolution of the first 2 Gyr in log-scale, which is dominated by the formation and subsequent evolution of a bar \fix{(after 5.2 Myr which corresponds to roughly 30 initial $t_{dyn}$, see eqn. \ref{eqn: tdyn})}, while the long term evolution (main panels) is slow and shows a gradual increase (decrease) in triaxiality (flattening).}
    \label{fig: axes ratios evolution}
\end{figure*}

The second moment tensor method (hereafter tensor method\footnote{Commonly also referred to as principal component analysis, see, for example, \citet{Freour+2026}.}), on the other hand, provides a more intrinsically motivated description by quantitatively characterising the distribution of stellar positions in three dimensions. By computing the eigenvalues and eigenvectors of the inertia (or mass) tensor, one can determine the principal axes and corresponding axis ratios of the stellar system. When applied to projected data, it yields an effective two-dimensional ellipticity; when applied in full three dimensions—as is possible with direct $N$-body simulations—it also gives access to the intrinsic cluster morphology, or its triaxiality. This method therefore facilitates a direct comparison between projected ellipticities and the underlying intrinsic structure. The triaxiality, $T$, is defined as

\begin{equation} \label{eqn: Triaxiality}
	T = \frac{a^2 - b^2}{a^2 - c^2}
\end{equation}
where $a,\,b,\,c$ are the major-, intermediate-, and minor axes of the system, respectively. These are in turn given by the square-root of the eigenvalues $\lambda_1,\lambda_2,\lambda_3$ of the second moment tensor $T_M$, which are obtained by diagonalising 
\begin{equation} \label{eqn: Tensor matrix}
T_M =
\begin{bmatrix}
I_{xx} & I_{xy} & I_{xz} \\
I_{yx} & I_{yy} & I_{yz} \\
I_{zx} & I_{zy} & I_{zz}
\end{bmatrix}.
\end{equation}
The entries of the moment tensor are calculated as $I_{ij} = \sum_k m \delta i_k \delta j_k$ where $m$ is the mass (which we set to be unity for all stars to give them equal weight), $\delta i_k$, $\delta j_k$ are the positions of stars with respect to the centre of mass, and $i,j$ are any of the cartesian coordinates $x,y,z$ for each star $k$. The eigenvectors $v_1,~v_2,~v_3$ contain additional information regarding the orientation of the three principal axes, and we obtain the azimuthal position angle $\phi = \mathrm{atan2}(v_1, v_2)$ and the polar position angle $\theta = \mathrm{cos}^{-1}(v_3)$. We confine the polar angle $\theta$ to the range $(0,180]\degree$, e.g. either aligned or anti-aligned with the z-axis, respectively.

\fix{We compare these two methods using the \texttt{ROLLIN'} simulation \texttt{1.5M-A-R4-10} in Appendix \ref{Contours or tensors}. For the remainder of the paper, we adopt the tensor-method to quantify morphology in our models, which allows us to more easily track temporal changes in their intrinsic axis ratios, triaxiality, and orientation.}

For simplicity, and in order to capture both the temporal and spatial variations for all our simulations, we restrict the measurements of morphology to three regions: (i) globally within the Jacobi radius, (ii) within the half-mass radius, and (iii) in-between the half-mass radius and the Jacobi radius. Hereafter, we also avoid measuring the projected ellipticity in the three cartesian planes, and instead utilise the three principal axes directly: the major axis $a$, intermediate axis $b$, and the minor axis $c$. For the majority of simulations, this will have little to no effect on the analysis, but it is relevant for \texttt{250k-A-R4-25-incl45} with inclined rotation axis, and can also become important at larger radii (e.g. close to the Jacobi radius) since the principal axes do not necessarily align to the direction of the cartesian coordinates, as, for example, is seen in the $xy$ projection in Fig. \ref{fig: tensor vs iso method}.

\section{Time evolution of morphology} \label{sect: Time evolution}

Paper I introduced our suite of self-consistent models of rotating GCs. The simulations presented in these works are characterised by a substantial amount of angular momentum, with the initial rotation axis aligned with the $z$-axis. As a consequence, the cluster morphologies show clear flattening perpendicular to the rotation axis, as expected for rotating systems. However, the detailed shape of the ellipticity profile, for example in model \texttt{1.5M-A-R4-10} (see Fig. \ref{fig: tensor vs iso method}), does not map in a simple one-to-one way onto the corresponding rotation profile derived for Fig 5 in Paper I \citep[a similar behaviour was also seen in, for example,][]{Bellini+2017}. In Paper I, we showed that our models exhibit differential rotation with a peak around the half-mass radius $r_{50\%}$ whereas the ellipticity tends to increase steadily toward the outskirts. This mismatch highlights the complex coupling between morphology and kinematics, and suggests that additional dynamical ingredients contribute to setting the global structure of GCs. In this section, we investigate the main physical processes that shape GC morphology, focusing in particular on the roles of internal rotation, velocity anisotropy, the external tidal field, and internal dynamics.

The first three panels of Fig. \ref{fig: axes ratios evolution} show the time evolution of \texttt{1.5M-A-R4-10} and its ellipticity $1-b/a$, $1-c/a$, and $1-c/b$, respectively, in three different regions of the GC: inside the Jacobi radius $r < r_j$ (dashed black line), inside the half-mass radius $r < r_{50\%}$ (dashed orange line), and in-between the half-mass radius and the Jacobi radius $r_{50\%} < r < r_j$ (dotted grey line). The fourth panel shows the triaxiality $T$ evolution of the GC (see eqn. \ref{eqn: Triaxiality}). The inset axes highlight the early morphological evolution, and show the first $2~\mathrm{Gyr}$ in log-scale. We include the evolution of $1-c/a$ and triaxiality for all remaining simulations in the \texttt{ROLLIN'} suite in Appendix \ref{Appendix: Full Triaxiality}, see Fig. \ref{fig: Full 1-c/a} and Fig. \ref{fig: Full Triaxiality}, respectively.

\begin{figure*}[htbp!]
    \centering
    \includegraphics[width=\linewidth]{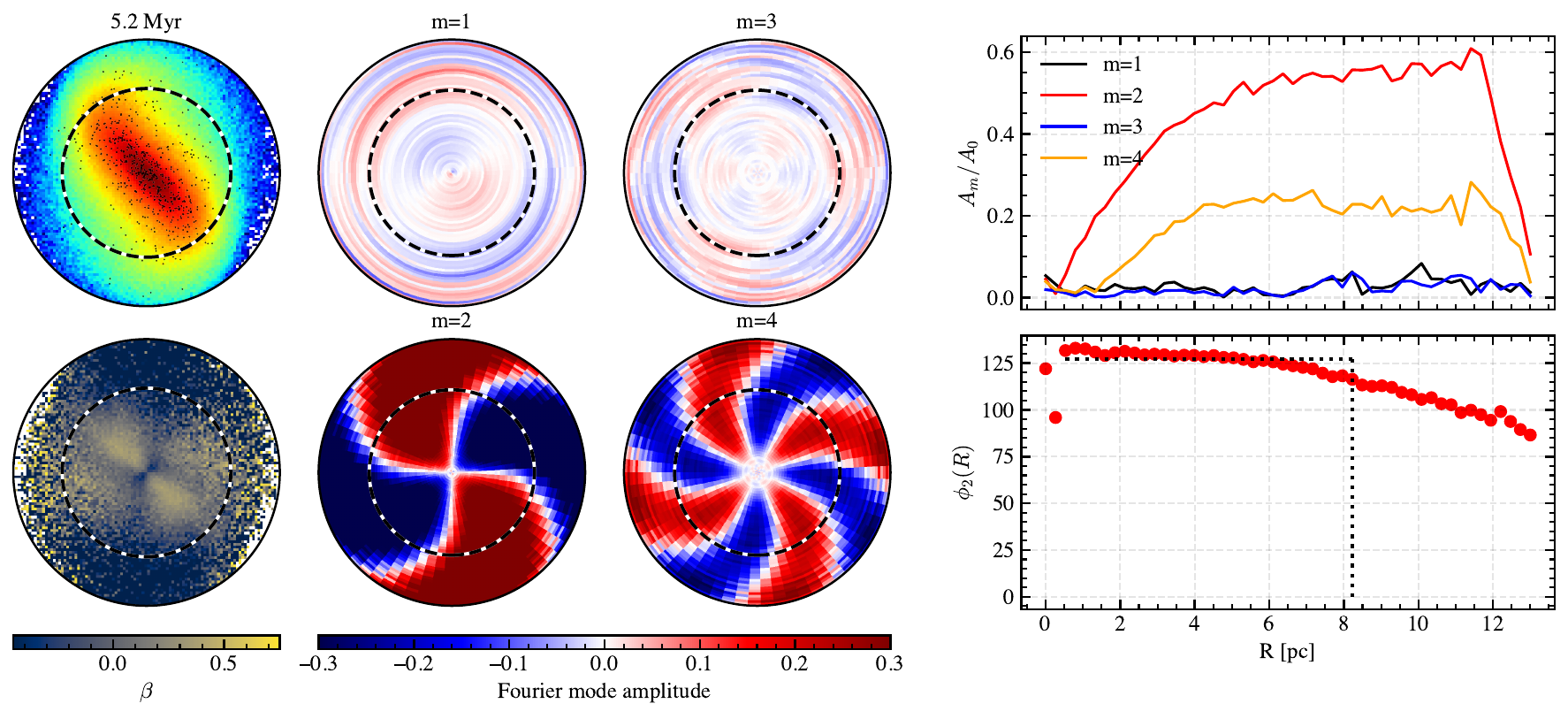} 
    \caption{Formation of a bar at $5.2~\mathrm{Myr}$ for \texttt{1.5M-A-R4-10}. The top left polar panel shows the number density and a bar-like structure viewed along the rotation axis. The bottom left panel shows the corresponding anisotropy parameter $\beta$, and the remaining panels show the four first Fourier modes (excluding $m=0$) normalised to $m=0$. The two panels to the right shows the amplitude of the different modes and the corresponding phase of $m=2$ as a function of radius. It appears that the $m=4$ mode behaves as an harmonic of the $m=2$ mode. The dashed lines in the various polar panels indicate the extent of the bar, which is also indicated in the bottom right panel along with the phase of the bar.}
    \label{fig: bar formation}
\end{figure*}

\subsection{Early evolution: dynamical instabilities and bar formation} \label{Sect: Bar formation}

The rotating models are initially set into an axisymmetric configuration around the $z$-axis. The ellipticity in the three main components for \texttt{1.5M-A-R4-10} are $1-b/a = 0$, $1-c/a = 1-c/b = 0.415$, respectively, with $T=0$. The subsequent evolution (see inset axes) is characterised by an increase in ellipticity, especially visible along $b/a$ within the first 10 Myr. Simultaneously, there is a strong increase (decrease) in $c/a$ ($c/b$); the GC is stretched and elongated along the major axis, resulting in a prolate shape ($T\lesssim 1$). This peak in ellipticity and triaxiality arises due to a dynamical instability leading to the formation of a bar-like structure. Fig. \ref{fig: bar formation} shows a snapshot of the simulation at 5.2 Myr (time of the spike) viewed along the rotation axis. The top left panel shows the number density (high in red, low in blue), and the black points represent black holes (BHs). A distinct bar-like structure can be seen spanning diagonally across the top left to bottom right. 

Previous numerical studies, such as \citet{Kamlah+2022}, have also found bar like-structures comprised of massive stars and stellar remnants in the early evolution of rotating GCs \citep[see also][]{Ernst+2007}. The formation of this bar-like structure appears to be the result of the combined effect of internal rotation and velocity anisotropy \citep[see][and \citealp{Tep+2025} which brought the analysis beyond spherical symmetry for the initial equilibria, finding a consistent behaviour]{Rozier+2019,Breen+2021}. In the bottom left panel of Fig. \ref{fig: bar formation}, we plot the anisotropy parameter, defined as $\beta = 1 - (\sigma_\theta^2 + \sigma_\phi^2)/ 2\sigma_r^2$ with ($\sigma_r,\sigma_\theta,\sigma_\phi$) representing velocity dispersion in the spherical coordinates, respectively, for the snapshot of \texttt{1.5M-A-R4-10}. $\beta > 0$ indicates radial anisotropy, and $\beta < 0$ indicates tangential anisotropy. 

\begin{figure}
    \centering
    \includegraphics[width=\linewidth]{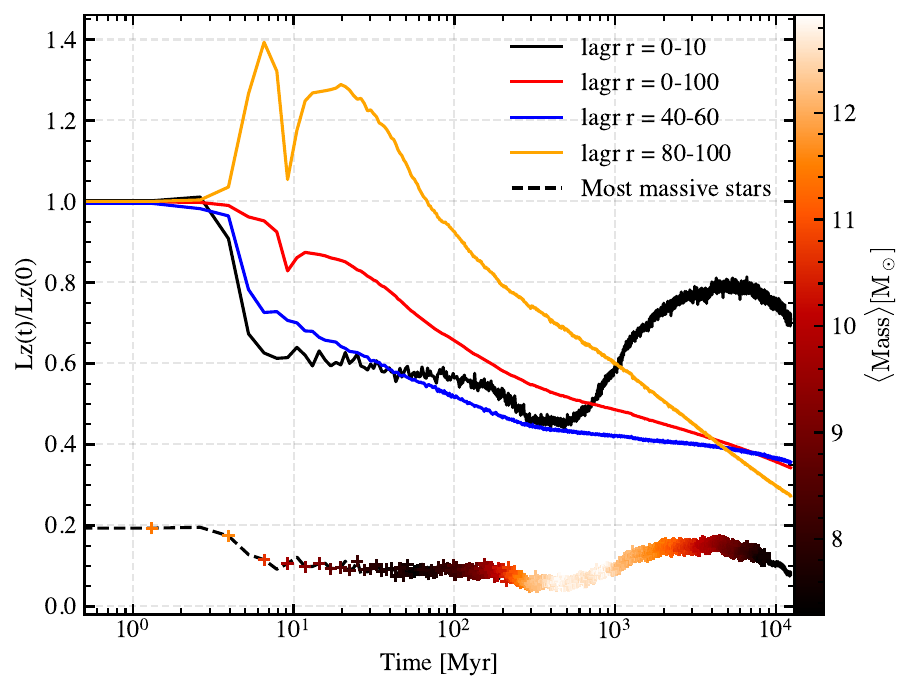}
    \caption{Angular momentum evolution in different Lagrangian shells. The early evolution is dominated by rapid loss of angular momentum within the inner regions, and a transport of angular momentum from the inner to the outer region. Conversely, the long term evolution is characterised by an increasing central angular momentum after the early core collapse (corresponding to the minimum within $r_{10\%}$).}
    \label{fig: Lz evolution}
\end{figure}

Our simulations are characterised by an initially isotropic central region, a radially anisotropic intermediate region, and a mildly tangentially anisotropic outer region. Within $5~\mathrm{Myr}$, the central and intermediate regions have become radially anisotropic. This is, in part, due to massive stars segregating efficiently in the central region as reported in Fig. 4 of Paper I. Most of these massive stars also form BHs through supernovae (SNe), providing a mechanism for rapid mass loss and rapid cluster expansion. Between $3-4~\mathrm{Myr}$ approximately 600 BHs have formed. This mechanism is also responsible for a loss of angular momentum, as already reported in Fig. 8 of Paper I, and the effects are also shown in Fig. \ref{fig: Lz evolution}. Each line shows the total angular momentum within different Lagrangian radial shells as a function of time, normalised to the respective shell's total initial angular momentum. In particular, the very central region (within the $10\%$ Lagrangian radius) begins to lose a considerable amount of angular momentum from $3-4~\mathrm{Myr}$. The pronounced drop in angular momentum at $5.2~\mathrm{Myr}$ corresponds to the snapshot in Fig. \ref{fig: bar formation}. Furthermore, the decrease (increase) of $L_z$ in the inner (outer) Lagrangian region, demonstrates an overall loss and transport of angular momentum from the inner parts of the GC to the outer parts. At this point in the simulation, the number of BHs within the Jacobi radius has increased to $\sim1500$, corresponding to a total mass loss of $\sim 5\%$ through SNe\footnote{This does not include mass loss due to BHs or neutron stars escaping from high natal kicks.}. 

The points along the dashed black line refers to the angular momentum for the 1000 most massive stars within the $10\%$ Lagrangian radius (they are, as such, normalised to the initial total angular momentum within that shell), and are coloured by their average mass. The most massive stars, which after $\gtrsim4~\mathrm{Myr}$ are dominated by BHs, almost exclusively populate a low angular momentum-, and radially anisotropic region.

We do not see the formation of a bar, nor the peak in ellipticity and triaxiality, for all simulations (see for example \texttt{250k-C-R2-10} in Fig. \ref{fig: Full Triaxiality}). Below, we make a short exploration of what additional conditions need to be present in order to trigger bar formation in the early evolution of GCs. We choose to describe the bar by defining its length $r_{bar}$, strength $S$, and phase $\phi$, following the methodology routinely used to characterise bars in disc galaxies \citep[see e.g.][]{Aguerri+1998, Kraljic+2012, Michea+2021}. This analysis is presented in Fig. \ref{fig: bar formation}, where the remaining four polar panels show the $m=1,2,3,4$ Fourier modes of the density distribution normalised to the $m=0$ mode, all with a common colour-bar. In the two panels on the right, we compute the amplitude of these modes with
\begin{equation}
A_m = \left|\displaystyle\sum_j m_j e^{i m\phi_j}\right| \times \left( \displaystyle\sum_j m_j \right)^{-1},
\end{equation}

and the corresponding phase ($m=2$) is determined by
\begin{equation}
\phi_{\rm bar} = \frac{1}{2} \arg\left(\sum_j m_j e^{i 2\phi_j}\right).
\end{equation}
Each $j$ labels a particle in the simulation, with $m_j$ being its mass, $\phi_j$ its azimuthal angle in cylindrical coordinates, and $e^{im\phi_j}$ its complex phase. The length of the bar can be determined by searching for a region of constant phase\footnote{We allow the phase to vary by at most $2\degree$ between consecutive points and terminate the bar when the cumulative phase variation exceeds $10\degree$.} \citep{Kraljic+2012, Michea+2021}. The phase and the length are indicated with the dashed horisontal and vertical black lines, respectively, in the bottom right panel of Fig. \ref{fig: bar formation}. The length, or extent, of the bar is also indicated by the dashed black line in each of the polar panels.

If a bar is identified, its strength is computed as
\begin{equation}
S = \frac{1}{r_{bar}} \int_{r_{start}}^{r_{end}} \frac{A_2}{A_0} dr,
\end{equation}
where $r_{bar}$ is the length of the bar, $r_{start}$ and $r_{end}$, are the inner and outer radius of the bar, respectively. We apply this procedure to all the snapshots of our suite of simulations.

\begin{figure*}[htbp!]
    \centering
    \includegraphics[width=\linewidth]{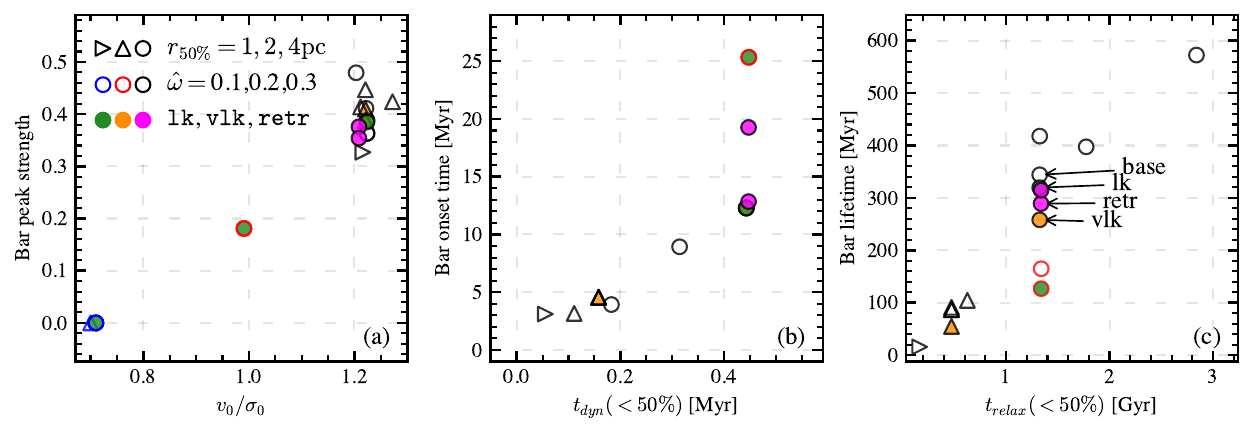}
    \caption{Relation between the bars' peak strength and \fix{initial} rotational support $v/\sigma$ (a), onset time and \fix{initial} dynamical timescale in Myr (b), and lifetime and \fix{initial} relaxation timescale in Gyr (c). We find no correlations between any other combination of these different quantities, see Fig. \ref{fig: bar vs ICs full}. Different symbols indicate models with different initial half-mass radii, different edge-colours indicate the initial rotation strength, and the different filling colours indicate models with any of \texttt{-lk}, \texttt{-retr}, or \texttt{-vlk}.}
    \label{fig: bar vs ICs}
\end{figure*}

Fig. \ref{fig: bar vs ICs} shows (i) how the bars' peak strength depends on the \fix{initial} peak of the velocity curve normalised by the central velocity dispersion ($v/\sigma$) which quantifies the relative importance of ordered rotation to random motions, and is a proxy for how rotationally supported the cluster is, (ii) how the bars' onset time of formation depends on the \fix{initial} dynamical timescale within the half-mass region, which is defined as
\begin{equation} \label{eqn: tdyn}
	t_{dyn} = \sqrt{\frac{2r_{50\%}^3}{GM}},
\end{equation}
where $r_{50\%}$ is the half-mass radius, $G$ is the gravitational constant, and $M$ is the total mass, and (iii) how the lifetime of the bars depends on the \fix{initial} relaxation time within the half-mass region \citep[see eqn. (2) in][]{Bianchini+2016}. We colour-code the simulations according to the structural properties of their initial conditions: different symbols correspond to different initial half-mass radii, and the edge colour indicates the initial rotation strength \citep[see Paper I and][]{VarriBertin2012}. Markers filled in green and orange correspond to the \texttt{-lk} and \texttt{-vlk} simulations, respectively (indicating different prescriptions for the natal kicks of stellar remnants), and those filled in fuchsia are retrograde simulations (see Section 2 in Paper I for a complete description).

Panel (a) shows that stronger bars are formed for the more rotationally supported systems, and that slow rotating models (i.e. those characterised by $\hat\omega=0.1$) do not form a bar at all (these are not present in the remaining panels). This is consistent with the expectation that a rotating stellar system may become unstable to bar formation above a critical threshold in total angular momentum \citep[][and various other global criteria; see the introduction of \citealp{Rozier+2019} for a summary of the relevant literature]{OstrikerPeebles1973}. The dynamical timescale instead determines how quickly the bar forms: models with shorter dynamical timescales form bars earlier, as seen in panel (b). It indicates that the instability is dynamical (collisionless): denser regions have shorter crossing times, so stars complete more orbits per unit physical time, allowing the unstable modes to grow faster. 

Panel (c) shows that the decay of the bar is quicker for shorter relaxation times (lifetime of the bar is thus longer for GCs with longer relaxation times). The bar decays because collisional relaxation progressively erases the anisotropic, low-($L_z$) orbital population that supports it, providing a dynamical heating mechanism. Since this is driven mainly by two-body relaxation, systems with longer ($t_{\rm relax}$), remain less collisional and redistribute angular momentum more slowly: the bar survives for longer. It is worth mentioning that, in the case of \texttt{1.5M-A-R4-10} (and a few other models, see Fig. \ref{fig: Full Triaxiality}), the bar is still intact (though weak) throughout the initial core collapse. Compare, for example, the time of minimum angular momentum within the $10\%$ Lagrangian shell in Fig. \ref{fig: Lz evolution} ($\lesssim 400\mathrm{Myr}$) and the time of fully restored axi-symmetry inside the half-mass radius in Fig. \ref{fig: axes ratios evolution} ($\lesssim 800\mathrm{Myr}$). A full comparisons between the bar properties (y-axis) and the rotational support and timescales (x-axis) is included in Appendix \ref{Appendix: Bar properties}.

We also see from Fig. \ref{fig: Full Triaxiality}, that models that do not form a bar (e.g. no sudden triaxial morphology early on), also do not undergo a pronounced core collapse, if at all (compare to e.g. Fig. A.1 in Paper I). This provides evidence that the physical ingredients responsible for the onset of the bar instability on a dynamical timescale, i.e. internal rotation and velocity anisotropy, also affect the overall timing and depth of the process of core collapse on a relaxation timescale \citep[for the role of anisotropy, see][for the role of rotation see \citealp{Ernst+2007,Tep+2024} among others]{Breen+2017,Pavlik+2024}.

Furthermore, Fig. \ref{fig: bar vs ICs} also shows that the onset and lifetime differs for different stellar evolution prescriptions. The \texttt{-lk} and \texttt{-vlk} models, e.g. simulations where the cluster retain more remnants (e.g. BHs) due to their lowered natal kicks, consistently form their bars later on (with the exception of \texttt{SIM-A-R2-25-vlk} and \texttt{SIM-B-R4-10-lk} in which the bar is formed at the same time as their normal counterparts) and decay at a faster rate. The latter is particularly clear for the \texttt{SIM-A-R4-25} simulation, which has both \texttt{lk}-, and \texttt{vlk}- counterparts (as well as a \texttt{retr}-counterpart). We indicate these with arrows in panel (i); with increasing lifetime we have the \texttt{vlk}- (\texttt{retr}-), \texttt{lk}-, and the base-version, respectively.

The dichotomy of the stellar evolution prescriptions can be the result of one or many different mechanisms. One plausible explanation is that with more remnants, there is more heating, and angular momentum is re-distributed more quickly. This could lead to a quicker decay of the $m=2$ mode \citep[see for example][for more details on bar evolution]{ContopoulosPapayannopoulos1980,Athanassoula2002,Athanassoula2003}. We leave a more detailed analysis for the bar and, in particular, its influence on core collapse and the evolution of BH populations in these simulations, for a future study.

After the formation of the bar, which corresponds to the spike in Fig. \ref{fig: axes ratios evolution} (see also Fig. \ref{fig: Full Triaxiality}), the flattening in $b/a$ as well as the triaxiality can be seen oscillating for the $r < r_j$ and $r_{50\%} < r < r_j$ regions while dampening over time. This was also seen in the outer parts of rotating GCs in \citet{Kamlah+2022}, during a period that they coin the "restoration of axisymmetry". We find that these oscillations are caused by the alignment of the bar and the tidal field; the ellipticity is maximum when the bar aligns with the direction of the tidal tails.

\begin{figure*}[htbp!]
    \centering
    \includegraphics[width=0.95\textwidth]{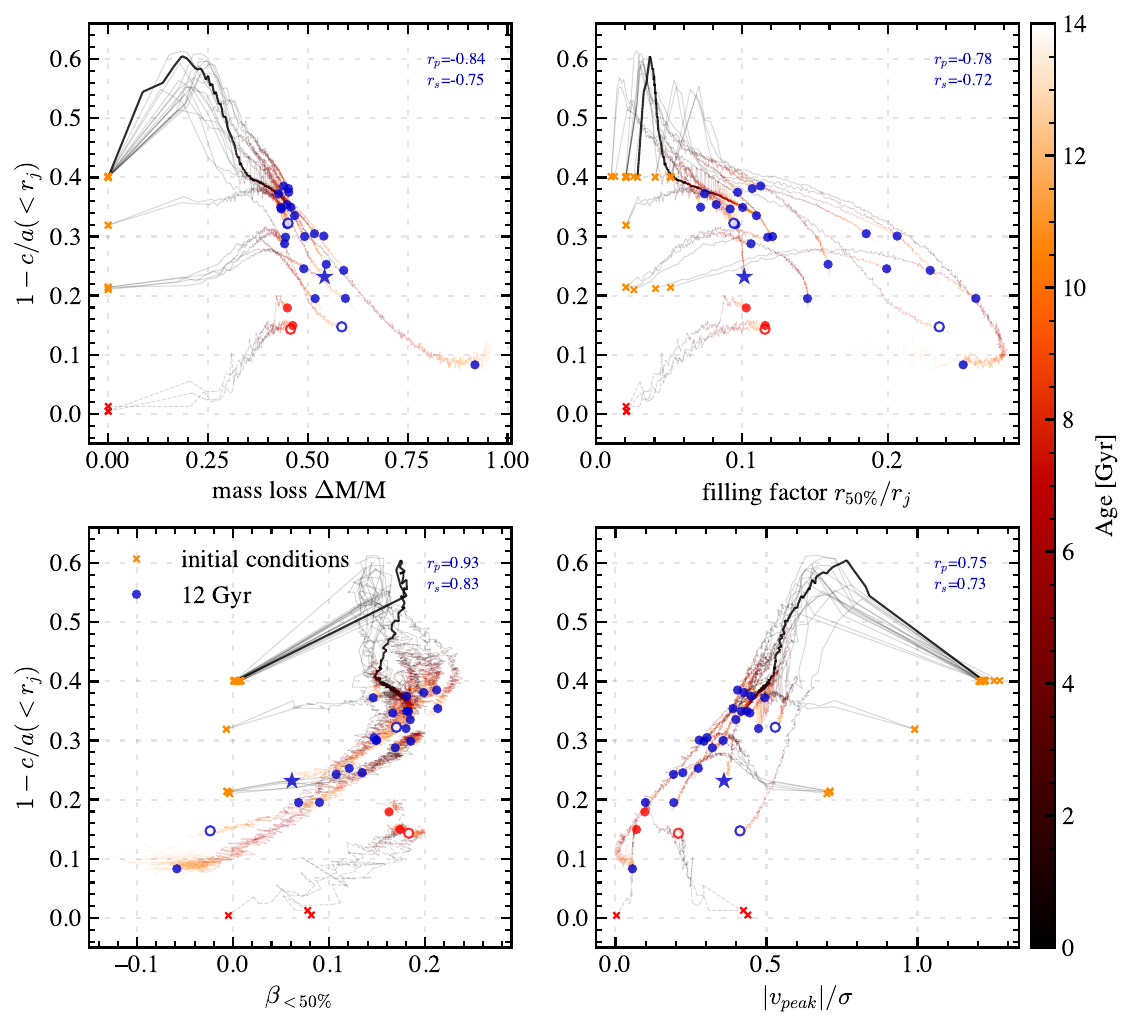}
    \caption{Correlations between the flattening ($c/a$) of all simulations within the Jacobi radius ($r_j$) and the corresponding mass loss, filling factor, anisotropy parameter ($\beta$, within the half-mass radius), and the rotational support $v_{peak}/\sigma$. Wilson models are highlighted in red, crosses indicate initial conditions (orange for standard models), lines indicate the evolutionary tracks (\fix{dashed for  spherical Wilson models}), and circles \fix{represent} the snapshots at 12 Gyr (blue for standard models). The open circle corresponds to models with an initial retrograde rotation, while the star corresponds to our densest model. The Pearson and Spearman coefficients when including the Wilson models are for mass loss $r_p=-0.64$, $r_s=-0.62$; filling factor $r_p=-0.59$, $r_s=-0.63$; anisotropy parameter $r_p=0.70$, $r_s=0.63$; and rotational support $r_p=0.80$, $r_s=0.78$.}
    \label{fig: exz vs stuff}
\end{figure*}

\begin{figure*}[htbp!]
    \centering
    \includegraphics[width=0.95\textwidth]{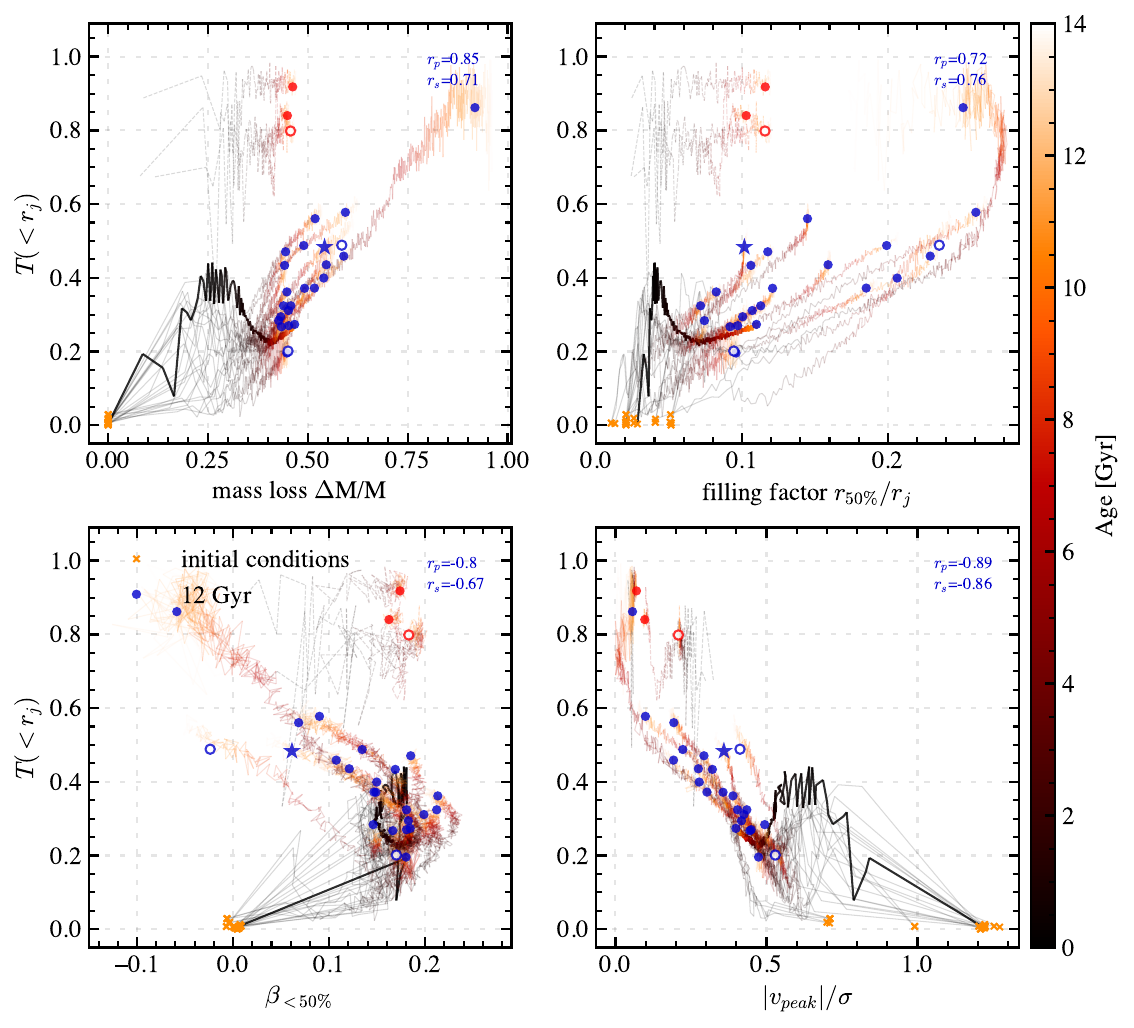}
    \caption{Same as Fig. \ref{fig: exz vs stuff} but for the triaxiality within the Jacobi radius. The Pearson and Spearman coefficients when including the Wilson models are for mass loss $r_p=0.46$, $r_s=0.56$; filling factor $r_p=0.37$, $r_s=0.65$; anisotropy parameter $r_p=-0.41$, $r_s=-0.47$; and rotational support $r_p=-0.89$, $r_s=-0.89$.}
    \label{fig: T vs stuff}
\end{figure*}

We interpret the rotation and mass segregation to be the main drivers for the onset of the unstable modes and the consequent formation of a bar-like structure. This interpretation is mostly consistent with what was presented by \citet{Kamlah+2022,Bissekenov+2025}. The authors ran their simulations both with and without stellar evolution, with a bar-like structure only appearing for the former. To confirm their findings, we ran two control simulations, using \texttt{250k-A-R4-25} and \texttt{250k-B-R4-25}, up to $900~\mathrm{Myr}$ and $300~\mathrm{Myr}$, respectively, with stellar evolution disabled; disabling stellar evolution produces an early evolution that is substantially different. The simulations without stellar evolution still form a bar-like structure at roughly the same time as when stellar evolution is enabled ($\sim 12\mathrm{Myr}$ and $\sim 25\mathrm{Myr}$, respectively), but they are sustained for much shorter: $\sim 50\mathrm{Myr}$ and $\sim 10\mathrm{Myr}$, respectively, which is roughly ten times as short compared to when stellar evolution is enabled ($\sim 340\mathrm{Myr}$ and $\sim 160\mathrm{Myr}$, respectively).

It suggests that the formation of the bars in our models are caused by dynamical instabilities determined by the initial amount of internal rotation and velocity anisotropy in our initial conditions. Stellar evolution, does, however play a role for the longevity of the bar. In the absence of stellar evolution, angular momentum is redistributed mainly through two-body relaxation and escapers, which is less efficient at depleting and maintaining a low angular momentum central region and radial anisotropy. With stellar evolution enabled, continuous mass- and angular momentum loss from massive stars, neutron stars, and BHs, maintains the inner regions close to the ROI threshold, allowing the bar to persist over several hundred Myr rather than being a short-lived transient. We include a side-by-side comparison with the two control models, with stellar evolution enabled and disabled, in Appendix \ref{Appendix: Bars} (see Fig. \ref{fig: StevNoStev}).

\subsection{Long-term evolution: correlations with dynamical and kinematic properties} 

The long-term morphological evolution of the models in our suite of simulations changes slowly (see, for example, the evolution in linear scale for Fig. \ref{fig: axes ratios evolution}), and it is also fairly similar regardless of whether the bar-like structure forms or not (see Fig. \ref{fig: Full Triaxiality}). We consider the early evolution (discussed in the section above) to be fully over once the bar-like structure is completely eroded by collisional effects; this coincides with the restoration of axisymmetry within the half-mass region (i.e. the ratio between the major and intermediate axes is roughly unity), as seen in Fig. \ref{fig: axes ratios evolution} (see also Fig. \ref{fig: Full Triaxiality}). In the case of \texttt{1.5M-A-R4-10} this occurs around $600-800~\mathrm{Myr}$, for other bar-forming models it ranges between $100-500~\mathrm{Myr}$, and in cases where no bar is formed, the clusters simply remain axisymmetric  (see Fig. \ref{fig: Full Triaxiality}). 

As was discussed in Paper I, stellar evolution dominates the mass loss early on, and can typically be attributed to a total loss of $30-40\%$ of the GC's initial mass. Typically, $30\%$ mass loss is reached within the first Gyr, after which the external tidal field and internal two-body interactions dominate the mass loss. In other words, throughout the long-term evolution of the models, we expect the tidal field and relaxation processes (as opposed to stellar evolution and dynamical instabilities) to shape the morphology.

From Fig. \ref{fig: axes ratios evolution} we find, in particular, a few general traits to be true across the standard rotational models (for the time being we exclude our spherical Wilson models in the analysis):
\begin{enumerate}
	\item[(i)] The flattening along the rotation axis decreases with time and the GC becomes slightly less axisymmetric (i.e. the GC becomes more triaxial) in the outer regions.
	\item[(ii)] The inner half-mass region remains oblate (i.e. the GC retains a symmetry axis around the rotation axis)
\end{enumerate}

In order to understand, and quantify, how different physical processes shape the GC morphologies, we explore how various structural and kinematical properties are correlated to the flattening of our models in Fig. \ref{fig: exz vs stuff}. More explicitly, we show the evolutionary tracks of the global flattening between the minor and major axis ($1-c/a$) with respect to the corresponding mass loss (top left), filling factor (top right), anisotropy within the half-mass radius (bottom left), and $v/\sigma$ parameter (bottom right). The axial ratio $c/a$ corresponds to the strongest flattening, and should simultaneously trace both the direction of the rotation axis ($c$) and the direction toward the galactic centre ($a$), e.g. the effects of the tidal field. 
The other quantities are chosen because they reflect the kinematics, the effects of the tidal field, or the general evolution of the cluster. In each panel, we also indicate the Pearson $r_p$ and Spearman $r_s$ coefficients to quantify any correlations\footnote{We have excluded the Wilson models in these calculations, but the corresponding coefficients for when they are included are indicated in the Fig. caption(s).}.

Initial conditions are indicated with crosses, and the properties at $12~\mathrm{Gyr}$ with circles. The spherical Wilson models are highlighted in red, the standard models in blue (grey lines), our densest model \texttt{250k-A-R1-10} as a star, and the open circles indicate clusters with an initial retrograde rotation with respect to their orbital angular momentum vector. The black line corresponds to the \texttt{1.5M-A-R4-10} model.

At time zero, as mentioned previously, the flattening of the clusters is fully determined self-consistently by the chosen initial equilibria, indicating that models with stronger rotation strengths are characterised by a higher degree of flattening. The stronger rotational models (with higher $v/\sigma$ in the bottom right panel), become highly flattened reaching an ellipticity upwards of $0.6$ at a mass-loss of around $15-25\%$ (these models are associated to the formation of the bar at this time). 
This roughly corresponds to the amount of mass loss associated with stellar evolution within the first $\gtrsim30~\mathrm{Myr}$ (see Fig. 9 in Paper I). The other models (with $\hat \omega < 0.3$) retain a roughly constant or moderate increase to their degree of flattening within $25\%$ mass loss. The ensuing evolution is then the same: increasing mass loss leads to less flattened systems. We relate this change in the intrinsic geometry to the perturbation induced by the external tidal field, which is more effective in the outer regions of the GCs, as we will discuss in the next section. There is also a small range in the flattening for similar mass loss after $12~\mathrm{Gyr}$: GCs that start more flattened also remain more flattened, given an otherwise similar dynamical history. The evolution of the flattening is also mirrored with respect to the filling factor. This is not surprising; Paper I demonstrated a strong correlation between mass loss and how tidally filling a GC is (see e.g. Fig. 9 in Paper I). Overall, we find a strong correlation between the flattening and the mass loss, as well as with the filling factor.

The two panels discussed above mainly show the effects driven by stellar evolution and the external tidal field. Paper I further demonstrated a relatively strong correlation between the anisotropy with both the mass-loss and filling factor, and for rotation strength with the total mass (see Figs. 10 \& 11 in Paper I). The remaining two panels in Fig. \ref{fig: exz vs stuff} (anisotropy and rotation) are meant to trace how the flattening correlates with the kinematical evolution of the GCs. We find a very strong correlation with anisotropy ($r_p=0.93$, $r_s=0.83$), and a relatively strong correlation with the level of rotational support ($r_p=0.75$, $r_s=0.73$). As GCs become less radially anisotropic, and lose rotational support, they become less flattened\footnote{Here we show the correlation with $v/\sigma$, but there is a slightly less strong correlation also for the peak rotation ($r_p=0.63$, $r_s=0.67$).}. 

For the quantities shown in Fig. \ref{fig: exz vs stuff}, the spherical Wilson models have distinct evolutionary tracks that are different to the standard models, and as a consequence they typically occupy a different parameter space. This is mainly due to an increasing flattening in their outer parts, as opposed to the decreasing flattening of the standard models. Interestingly, however, the relation between their rotational support and flattening at $12~\mathrm{Gyr}$ follows the same overall trend as for the standard models. We instead find that clusters with initial retrograde rotation deviate systematically from the overall relation. When these retrograde models are excluded (but non-retrograde Wilson models are included), the correlation between flattening and rotational support becomes very strong: the Pearson and Spearman coefficients are $r_p=0.92$ and $r_s=0.88$, respectively. This indicates that, despite using very different initial rotational prescriptions, a GC’s flattening seems to naturally align with how rotationally supported it is. Retrograde clusters lose less angular momentum due to the preferential stripping of prograde orbits, so they remain more rotationally supported. However, for these retrograde models, the higher rotational support does not lead to enhanced flattening. We see a similar behaviour for our densest model, \texttt{250k-A-R1-10} (marked as a star), and as such, it can be considered an outlier as well. Its higher density and shorter relaxation time drives it towards isotropy and sphericity while preserving rotational support.

We include similar figures to Fig. \ref{fig: exz vs stuff} in Appendix \ref{Appendix: Remaining Correlations}, but for $b/a$ (see Fig. \ref{fig: exy vs stuff}) and $c/b$ (see Fig. \ref{fig: eyz vs stuff}). For the latter we find very similar tracks as for the $c/a$ projection, but with a consistently less pronounced flattening, per definition. The $b/a$ projection shows no clear correlations with any of these quantities. 

Fig. \ref{fig: T vs stuff} shows these same evolutionary tracks, but with respect to the triaxiality of the simulations. Because the Wilson models are spherical at $t=0$, they are not defined\footnote{For all points we require that at least one of the principal axes has an absolute difference of $\geq0.05$ to the remaining axes.} in $T$ since $a\approx b\approx c$, while the remaining simulations all start oblate ($T\sim0$). The overall interpretation for the triaxiality remains similar to the flattening: as the clusters lose mass and angular momentum, the rotational support diminishes, they become less radially anisotropic, and the clusters become more triaxial, also in response to the perturbation imparted by the external tidal field. The triaxiality is also strongly anti-correlated with the rotational support; the more angular momentum that is lost, the more triaxial a cluster becomes. This implies that the effects of the tidal field become more visible for clusters that have less rotational support. Finally, the sphericals Wilson models remain extremely prolate ($T\sim1$) throughout their evolution. All of these correlations, and anti-correlations, are seen also within the half-mass radius, see Fig. \ref{fig: T50 vs stuff}.

\begin{figure*}[htbp!]
    \sidecaption
    \includegraphics[width=12cm, trim=0cm 0cm 8.5cm 0cm, clip]{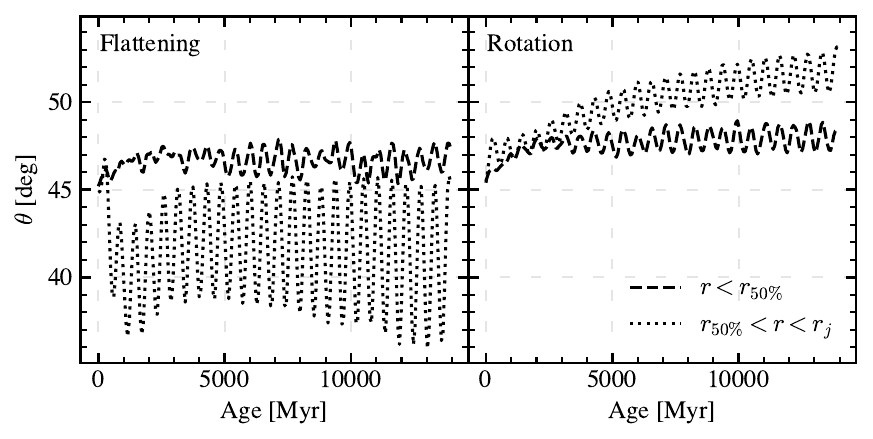}
    \caption{Time evolution of the position angle for the minor axis (first panel) and the rotation axis (second panel) for \texttt{250k-A-R4-25-incl45}. The two lines correspond to the region within (dashed) and outside (dotted) the half-mass radius.
    }
    \label{fig: polar angles}
\end{figure*}

\subsection{Flattening and triaxiality caused by the tidal field} \label{sect: Tidal field}

Triaxiality is a natural outcome of the presence of an external tidal field \citep[see e.g.][]{VarriBertin2009,ArditiVarri2026}, and the tidal field should influence the outer parts of a cluster more significantly, with the inner parts mainly being shaped by internal processes. For this reason, we include the same evolutionary tracks shown in Figs. \ref{fig: exy vs stuff} ($b/a$), \ref{fig: exz vs stuff} ($c/a$), \ref{fig: eyz vs stuff} ($c/b$), and \ref{fig: T vs stuff} ($T$), but for the inner half-mass region in Figs. \ref{fig: exy50 vs stuff}, \ref{fig: exz50 vs stuff}, \ref{fig: eyz50 vs stuff}, and \ref{fig: T50 vs stuff}, respectively. A few points from these figures underscore the influence of the tidal field on the flattening and morphology of the GCs:
\begin{enumerate}
	\item[(i)] The half-mass triaxiality is always smaller than the corresponding global triaxiality (Figs. \ref{fig: T vs stuff} and \ref{fig: T50 vs stuff})
	\item[(ii)] The filling factor, in particular, is strongly correlated with the triaxiality within the half-mass radius (Fig. \ref{fig: T50 vs stuff})
	\item[(iii)] Ellipticity along $b/a$ increases with increasing filling factor, most notable within the half-mass region, but is not correlated with the kinematic quantities as the angular momentum vector is perpendicular to the $ab$-plane (Fig. \ref{fig: exy50 vs stuff})
	\item[(iv)] The standard models with a modest mass loss ($\lesssim 50\%$) remain fairly oblate (Figs. \ref{fig: T vs stuff} and \ref{fig: T50 vs stuff})
	\item[(v)] For more significant mass loss, triaxiality changes both quickly and drastically (Figs. \ref{fig: T vs stuff} and \ref{fig: T50 vs stuff}), indicating that the tidal field had time to act effectively
	\item[(vi)] The Wilson models only show a significant flattening for the outer regions between $b/a$ and $c/a$ (Fig. \ref{fig: exz vs stuff}), suggesting that $a$, being in the direction of the Galactic centre, is influenced by the tidal field
	\item[(vii)] The Wilson models remain roughly spherical within the half-mass radius, with $T$ displaying a statistical scatter (Fig. \ref{fig: T50 vs stuff}) and $e\sim0$ for the different projections (Figs. \ref{fig: exy50 vs stuff}-\ref{fig: exz50 vs stuff})
	\item[(viii)] The three Wilson models are prograde, retrograde, or non-rotating with respect to the GC orbital angular momentum, and none is initially flattened along the minor axis. Any flattening that develops can therefore be attributed, at least in part for the initially rotating models, to the external tidal field. As shown in Fig. \ref{fig: exz vs stuff}, all three models have developed a clear non-zero flattening after $12$ Gyr, and the initially non-rotating model has developed a small but non-zero degree of rotational support. This is consistent with tidal-field induced rotation because of preferential stripping of prograde orbits \citep[i.e.][]{Tiongco+18}.
\end{enumerate}

The first point shows that the tidal field affects the outer parts more strongly, as expected. Clusters that are more tidally filling naturally resemble the shape of the Roche lobe, thus becoming more triaxial, as demonstrated by points (ii-v). Because morphological observations typically are confined to a few times the inner half-mass region, it is useful to note that the triaxiality still correlates with the other quantities, and, in particular, with the filling factor. The last three points (vi-viii), in particular, highlights that the tidal field is capable of imprinting a significant non-zero flattening, as well as rotation.

\section{Special case: inclined rotation angle} \label{Sect: 45 inclination}
In this work we included a new model, \texttt{250k-A-R4-25-incl45}, which explores a more general configuration by tilting the rotation axis by \(45\degree\) relative to the orbital angular momentum vector. This introduces a more complex coupling between internal rotation, morphology, and the external tidal field. As shown in Fig. \ref{fig: Full Triaxiality}, the model develops a bar-like structure, and its overall triaxiality evolution (and thus its flattening) follows the same qualitative trend as the other models. However, its evolutionary tracks in our correlation figures, e.g. Figs. \ref{fig: exz vs stuff} and \ref{fig: T vs stuff} (see also Figs. \ref{fig: exy vs stuff}-\ref{fig: T50 vs stuff}), further show that an initially inclined rotation axis leads to a similar evolution to that of the initially retrograde models. In particular, for the correlation between flattening and rotational support in Fig. \ref{fig: exz vs stuff} (bottom right panel), the inclined model is located just to the left of our most rotationally supported retrograde model (\texttt{250k-A-R4-25-retr}), indicating that it retains more rotational support than the prograde models, while maintaining a similar level of flattening. Furthermore, it stands out among all other models due to a distinct oscillating morphology that is not seen for aligned or anti-aligned rotation axes (Fig. \ref{fig: Full Triaxiality}).

\citet{Tiongco+18,Tiongco+2022} explored different position angles for rotating GCs with equal-mass particles. They found, in particular, that the rotation axis precesses as a consequence of torques from the tidal field of the host galaxy, and that there is an evolution towards a re-alignment between the rotation axis and the orbital angular momentum vector (z-axis). This mechanism was also found to have a radial dependence; the inner regions are dominated by intrinsic rotation and typically remain aligned with the initial inclination of the rotation axis, whereas the outer regions approximately rotates as a solid-body about the z-axis, possibly also imprinting counter-rotation. The orientation of the rotation axis was also found to mis-align with the orientation of the morphological minor axis, with the effects being more pronounced at larger radii.

The oscillations in morphology for \texttt{250k-A-R4-25-incl45} occur at a periodicity equal to half the orbital period. As such, the axial ratios depend on the orbital phase of the cluster. We find that our initially inclined model reaches local maxima in all axial ratios when the planar component of the minor axis is perpendicular to the direction of the tidal field (-$\hat x$); the tidal field mainly stretches the major axis (this corresponds to the phase in which the angular momentum component in the direction of the tidal field is zero). Conversely, at local minima in axial ratios, the planar component of the minor axis is aligned with the direction of the tidal field and is therefore the main axis to be stretched. These oscillations in morphology are not seen within the half-mass radius (see Fig. \ref{fig: Full Triaxiality}), and therefore reflect the outer morphology of the GC.

Although the axial ratio within the half-mass region does not oscillate, the corresponding position angle of the minor axis does. We define an inclination of $0\degree$ to be aligned with the orbital angular momentum vector (z-axis), and $180\degree$ to be anti-aligned (as such $90\degree$ is perpendicular to the orbital angular momentum vector, i.e. lying in the orbital plane). The first panel of Fig. \ref{fig: polar angles} shows the position angle $\theta$ of the minor axis as a function of time, within the half-mass radius (dashed line) and outside the half-mass radius (dotted line). These oscillations also occur with a periodicity of half the orbital period. At maximum flattening, when the minor axis is perpendicular to the tidal field and reaches a local minimum, the position angle of the minor axis reaches a local maximum (i.e. pointing further away from the orbital angular momentum vector). The oscillations for the inner and outer regions of the cluster coincide, but a mis-alignment between the inner and outer parts is visible; the mean position angle of the minor axis in the outer part is more aligned with the z-axis than the inner part, which in turn, follows more closely to the rotation axis. This means that the outer part is more elongated in the orbital plane due to tidal effects, explaining the smaller position angle.

The second panel shows the same evolution but for the position angle of the rotation axis within and outside the half-mass radius. Interestingly, we find the outer region to align in the opposite direction of the minor axis, i.e. less aligned with the z-axis. As such, we do not find a re-alignment of the rotation axis with the orbital angular momentum vector, contrary to \citet{Tiongco+18}. Their simulations consists of considerably fewer stars than in our models, they are also are run over more initial relaxation times than our simulations, and their models also lose considerably more mass ($75\%$). This discrepancy may therefore be a consequence of different dynamical ages. The mis-alignment of the rotation axis can be explained as a consequence of the presence of the tidal field. In particular, the preferential stripping of stars in prograde orbits \citep{Tiongco+18} induces a counter-rotation (in the reference frame of the GC orbit). The counter-rotation is mainly planar and anti-aligned with the z-axis, resulting in the outer (and global) rotation axis to align towards the orbital plane. As a consequence, the morphological minor axis and the kinematic rotation axis become mis-aligned \citep[consistent with the findings of][]{Tiongco+2022}.

\section{Comparisons with observations} \label{sect: Observations}
 
\begin{figure*}[htbp!]
    \centering
    \includegraphics[width=0.95\linewidth]{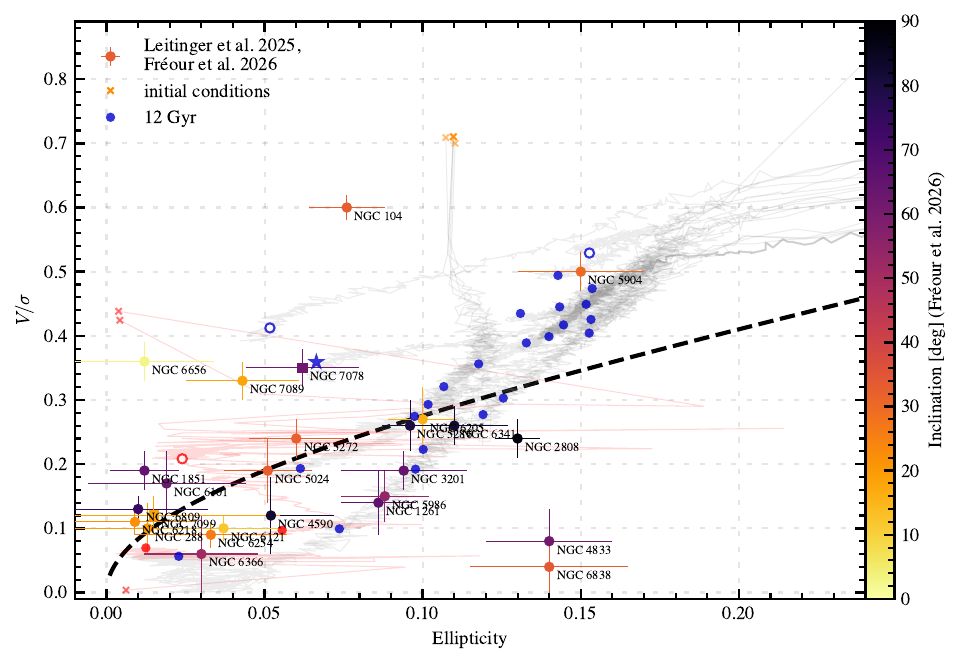}
    \caption{Comparison between measured $v/\sigma$ and ellipticity values for 29 Galactic GCs \citep[values taken from Tables C.1 and A.1 in][respectively, shown as coloured points based on their inclination angle]{Leitinger+2025,Freour+2026}, the \texttt{ROLLIN'} suite evolutionary tracks within the half-mass radius for $1-c/a$, and the supposed relation between ellipticity and $V/\sigma$ for an edge on view of a fully isotropic oblate rotator \citep[\fix{black} dashed line, see eqn. (6) in][]{Freour+2026}. As for Figs. \ref{fig: exz vs stuff} \& \ref{fig: T vs stuff}, the corresponding positions at 12 Gyr for our models are shown as blue (red) points for the standard models (spherical Wilson models), and open circles indicate initially retrograde models.}
    \label{fig: Galactic GCs}
\end{figure*}

The evolutionary tracks of our simulations reveal a tight connection between rotational support and intrinsic flattening, but also identify clear outliers, particularly among the retrograde models and the densest cluster in our suite (\texttt{250k-A-R1-10}). Building on this, we compare our models to observations of 29 Galactic GCs by plotting ellipticity against the total \(V/\sigma\) ratio in Fig. \ref{fig: Galactic GCs}. We use ellipticities from \citet{Freour+2026} and the three-dimensional \(V/\sigma\) measurements from \citet{Leitinger+2025}. The observed clusters span a similar range in the total rotational support to our models (both rotation measurements should reflect intrinsic rotational support), and more or less for the flattening as well (our models show $1-c/a$ within the half-mass radius), although their values are systematically lower. This is due to the fact that ellipticity measurements are projected quantities, while our measurements are intrinsic. We have not corrected the observed values of ellipticity for projection effects, and as such they should be regarded as lower limits. The commonly used reference for de-projecting ellipticity, and for correlating rotational support with ellipticity, is the isotropic oblate rotator \citep[black dashed line; e.g.][eqn. 6]{Freour+2026}, which has been used extensively for elliptical galaxies \citep[see e.g.][]{Cappellari+2007,Emsellem+2011,Brough+2017}. 

Not surprisingly, our simulations occupy a distinct band in the \(V/\sigma\)–ellipticity plane, following a relation that differs noticeably from the isotropic oblate rotator expectation, and that reflects the complex evolution driven by internal and external processes. We report the corresponding values\footnote{The full tracks are available here: \url{https://github.com/arnmarklund/ROLLIN.git}} for ellipticity and triaxiality of our models at $12\mathrm{Gyr}$ in Table \ref{tab: Morphology} to facilitate future use by the community. Many Galactic GCs fall within or close to this band, implying that clusters appearing as outliers to the isotropic oblate rotator are not necessarily inconsistent with more realistic rotational models. This is particularly evident for four systems—NGC 4590, NGC 5286, NGC 6341, and NGC 2808—which are observed close to edge-on, and for which projection effects should be minimal. Their measured ellipticities should therefore be close to their intrinsic values, and indeed they lie along, or very near, the locus traced by our simulations. 

A small number of clusters remain genuine outliers. The most prominent case is NGC 104 (47 Tuc), which exhibits the highest rotational support in the sample but only modest observed flattening. Dynamical models that fit the full three-dimensional kinematics of 47 Tuc indicate a peak intrinsic ellipticity of up to \(\epsilon \sim 0.45\) at radii of order \(6\,r_{50\%}\) \citep{Bellini+2017}, which would bring it much closer to our model's flattening in the outer regions, see Fig. \ref{fig: exz vs stuff}. Our densest simulation, though considerably less concentrated than 47 Tuc, similarly shows that a cluster can retain substantial rotational support while appearing nearly round in projection. Denser GCs are likely to isotropise more quickly in their central regions owing to shorter relaxation times, likely leading to a more circular appearance; this is also the case for \texttt{250k-A-R1-10} as seen from Fig. \ref{fig: exz vs stuff}. Consistent with this are the three core-collapsed Galactic GCs in this sample \citep[shown as squares][]{Trager+1995}, which also appear circular. One of them, NGC 7099, shows little to no rotational support, and cannot be considered as an outlier. NGC 7078, however, combines a nearly round shape with significant rotational support and lies close to the track of our densest model.

Other outliers can likely be explained by projection and environmental effects. NGC 6656, for example, has relatively weak flattening but considerable rotation; it is believed to be viewed close to face-on, so its intrinsic ellipticity could be substantially higher and more in line with our simulations. The perhaps most puzzling cases, also discussed extensively in \citet{Freour+2026}, are NGC 6838 and NGC 4833, which show significant flattening but almost no detectable rotation \citep{Bianchini+2018,Leitinger+2025}. NGC 6838 has a particularly low mass (\(\sim 5\times10^{4}\,{\rm M_{\odot}}\); \citealt{BaumgardtHilker2018}) suggesting it could have suffered substantial mass loss, which, based on our models should lead to pronounced triaxiality even within the half-mass radius (see, for example, \texttt{250k-A-R2-5}). Both NGC 6838 and NGC 4833 are thought to have undergone recent disc crossings and are expected to lose further mass in future passages \citep{Ferrone+2023,Pancino+2024}, processes that may enhance outer distortions. Although more significant mass loss generally leads to a smaller flattening according to our simulations, our models do not consider any time variability in the tidal perturbation as due to eccentric orbits, non-spherical or multi-component external potentials, and, in the more extreme instance, even disc crossings. Such mechanisms seem plausible for explaining the positions of these two clusters, particularly if future morphological studies were to uncover any evidence of isophotal twisting. However, their inclinations are also poorly constrained, making it difficult to determine how far they might move in the \(V/\sigma\)–ellipticity plane if deprojected. 

It should also be noted, that the observational ellipticities used here are global values measured over radial ranges extending from \(\sim 2\) to \(19\,r_{50\%}\) \citep{Freour+2026}, whereas our models shown in this plot are confined to the inner half-mass region. When limiting the radial range to $3r_{50\%}$, \citet{Freour+2026} did not find a significant deviation for the measured flattening, but the radial range is still well outside the half-mass radius. This mismatch in radial coverage further complicates a strict one-to-one comparison but does not erase the overall agreement between the simulations and the bulk of the Galactic GC population.

\section{Conclusions} \label{Conclusions}

In Paper I we introduced our \texttt{ROLLIN’} suite of direct $N$-body simulations of initially rotating GCs. We explored, in particular, the impact of internal rotation in GCs on kinematic and dynamical properties for a wide range of initial conditions. In this study, we extend the suite with two new simulations (\texttt{250k-A-R4-25-incl45}, and \texttt{250k-W6-R4-25-norot}), with the goal of characterising their morphology, quantified by ellipticity and triaxiality, in tandem with other dynamical and structural properties. More specifically, we have considered two common methodologies used in the literature for deriving ellipticities: (i) iso-density contours, and (ii) the second order tensor method. Although there was a qualitative agreement between the two methods, we showed that they may also lead to non-negligible differences for the flattening at different radii, especially in the inner-most and outer-most regions. We chose to proceed with the tensor method due to its robustness, and because it provides a full three-dimensional and intrinsic description of the clusters' morphology.

Our work shows that GC morphologies are shaped by multi-scale dynamical processes, producing a rich set of features observable in GCs. The main drivers of GC morphological evolution are internal rotation, external tidal interactions, dynamical instabilities, two-body relaxation. In particular:

\begin{enumerate}
\item On a short timescale, stronger initial rotation and velocity anisotropy can lead to the development of unstable modes within the first few tens of Myr, and the consequent formation of a short-lived bar. This dynamical behaviour appears to be consistent with the extension of the classic radial orbit instability to the case of rotating systems \citep[see][]{Rozier+2019}. Stellar evolution plays a vital role, particularly for sustaining the bar for longer periods of time ($\approx$ a few $100~\mathrm{Myr}$), since it provides a mechanism for rapid mass loss (and loss of angular momentum) through SNe that maintains the instability. The subsequent evolution of the bar and the GC leads to a very characteristic morphology for our models:
\begin{enumerate}
	\item[(i)] The ellipticity of the projected GC viewed along the rotation axis reaches a maximum $e\lesssim0.4$, and a prolate morphology $T\lesssim1$.
	\item[(ii)] The inner half-mass region and the outer half-mass regions show two contrasting morphologies; the former (being dominated by a bar) remains prolate for up to a few hundred Myr, and later evolves towards a restored axisymmetry due to two-body relaxation (i.e. erosion of the bar), while the latter becomes triaxial due to the tidal field.
	\item[(iii)] The initial core collapse appears to only occur for the models that form a bar, implying a relation with the processes that sustains the bar. In particular, initial core collapse only appear in models that are able to quickly segregate BHs and massive stars in their centres, causing a significant redistribution of angular momentum. The connection between the bar, initial core collapse, and the overall evolution of BHs are left for a future study.
\end{enumerate}

\item The long-term morphological evolution for all models, regardless of whether they form a bar or not, remain quite similar. We showed that the maximum intrinsic flattening of a GC is strongly correlated with certain physical conditions. In particular, the flattening decreases with increasing mass loss, increasing filling factor, lower radial anisotropy, and decreasing rotational support, respectively. Similarly, the triaxiality of a cluster is also correlated to these quantities: higher mass loss, increasing filling factor, reduced radial anisotropy, and a diminishing rotational support all lead to a more pronounced triaxiality. The last point, in particular, highlights that the effects of the tidal field are more apparent for clusters that have less rotational support.

\item We also included three initially spherical \citet{Wilson1975} models to explore other initial conditions, in which two are rotating (one prograde \& one retrograde) and one which does not rotate. They show a distinct evolution, different to that of our standard, self-consistently determined rotating equilibria models. Specifically, they develop a pronounced maximum flattening $e\lesssim 0.2$ after $12~\mathrm{Gyr}$, comparable to the corresponding values for some of the initially flattened models. Because this is the case for all three models, and in particular for the non-rotating model, it implies that the flattening of a GC can be shaped by the tidal field: this can in part help explain why some flattened Galactic GCs do not show clear signs of rotation \citep[see for example][]{Freour+2026}.

\item The flattening along the major and minor axis, also naturally correlates with how rotationally supported a GC is. Deviations from this correlation, could be caused either by initially denser clusters that retain rotational support while becoming more isotropic, or by GCs with an initially inclined or retrograde rotation with respect to its orbital angular momentum vector, which retain more rotation due to the preferential stripping of prograde orbits, yet show similar levels of flattening. As such, it can also, in part, help explain how some rotationally supported GCs show a less pronounced flattening than expected based on their $v/\sigma$.

\end{enumerate}

The qualitative agreement between the dynamical, kinematical (Paper I), and now morphological properties of our models and Galactic GCs demonstrates that the \texttt{ROLLIN’} suite provides a valuable framework for interpreting both existing and upcoming observations. Nevertheless, as noted in Paper I, our models remain idealised in several respects: circular orbits in a point-mass potential, single stellar populations, and no primordial binaries. Eccentric orbits, disc crossings, and multi-component potentials (bulge + disc + halo) enhance mass loss, tidal stripping, and outer heating through pericentre shocks and impulsive energy injection \citep{GnedinOstriker1997,BaumgardtMakino2003,Renaud+2011,Webb+2013,Cai+2016}. These processes likely influence GC morphology directly or alter how dynamical and kinematical quantities correlate with it. They should, however, preferentially affect the outer regions; driving isotropy, reducing rotation, and modifying triaxiality, including isophotal twisting. Consequently, projection effects and limited observational apertures may mask these changes.

\fix{In our simulations we find that the evolution is governed by the coupled action of rotation, stellar evolution, and the external tidal field, with their relative importance varying across different phases. As shown by previous work \citep{Lamers+2010}, disentangling such effects is not straightforward, in particular because they are interdependent. For example, stellar evolution alters the cluster structure and mass which in turn changes how the tidal field acts on the system. A more detailed separation of these contributions will require future models with self-consistent, time-dependent tides and stellar evolution, which we leave for future work.}

Initially axisymmetric (or spherical) GCs may also not mirror realistic initial conditions during the formation, or assembly, of GCs. We will in a future study extend the current suite of the \texttt{ROLLIN'} simulations with additional models having less idealised initial conditions \citep[e.g., see][]{Vesperini+2014,Lahen+2020}. These are intended to reflect complex interplays between gas and stars during cluster formation, and will explore how spatially non-isotropic initial conditions — specifically, subcluster configurations — can lead to rotation and different morphologies through violent relaxation and angular momentum transfer.

In summary, we have provided a physical interpretation of how observed morphologies correlate with kinematical and dynamical properties of GCs. Our results, highlight the importance of developing reliable and consistent methods for measuring GC morphologies, as this quantity provides useful information on GCs' formation and subsequent evolution. Current and forthcoming large-scale deep photometric surveys calls for state-of-the-art techniques (e.g. deep learning algorithms, Marklund et al., in prep.) capable of retrieving accurate measurements of ellipticity, while simultaneously being scalable and computationally efficient, further enabling applications to vast and extensive datasets.

\section*{Data Availability} \label{DataAvailability}

\fix{Evolutionary tracks from the \texttt{ROLLIN’} suite are available here: \url{https://github.com/arnmarklund/ROLLIN.git}. Additional simulation outputs are available from the corresponding author upon reasonable request.}

\begin{acknowledgements}
We would like to thank our referee, Christophe Pichon, for the useful comments that helped in improving the manuscript. The authors thank L. Fréour for useful discussions. This project was provided with computing resources by GENCI at IDRIS thanks to the following time allocations on the supercomputer Jean Zay (V100 partition): Grand Challenge-101470, A10-A0100412451, and A13-A0130412451. The authors would also like to acknowledge the High Performance Computing Center of the University of Strasbourg for supporting this work by providing access to computing resources. Part of the computing resources were funded by the Equipex Equip@Meso project (Programme Investissements d'Avenir) and the CPER Alsacalcul/Big Data. AM, PB, and GP acknowledge financial support by the IdEx framework of the University of Strasbourg. ALV acknowledges support from a UKRI Future Leaders Fellowship (MR/S018859/1; MR/X011097/1). This work is partially supported by the grant GALBAR ANR-25-CE31-4684 of the French Agence Nationale de la Recherche.
\end{acknowledgements}

\bibliographystyle{aa}
\bibliography{references}

\begin{appendix}

\section{\fix{Contours or tensors}} \label{Contours or tensors}

\begin{figure*}[htbp!]
    \centering
    \includegraphics[width=\linewidth]{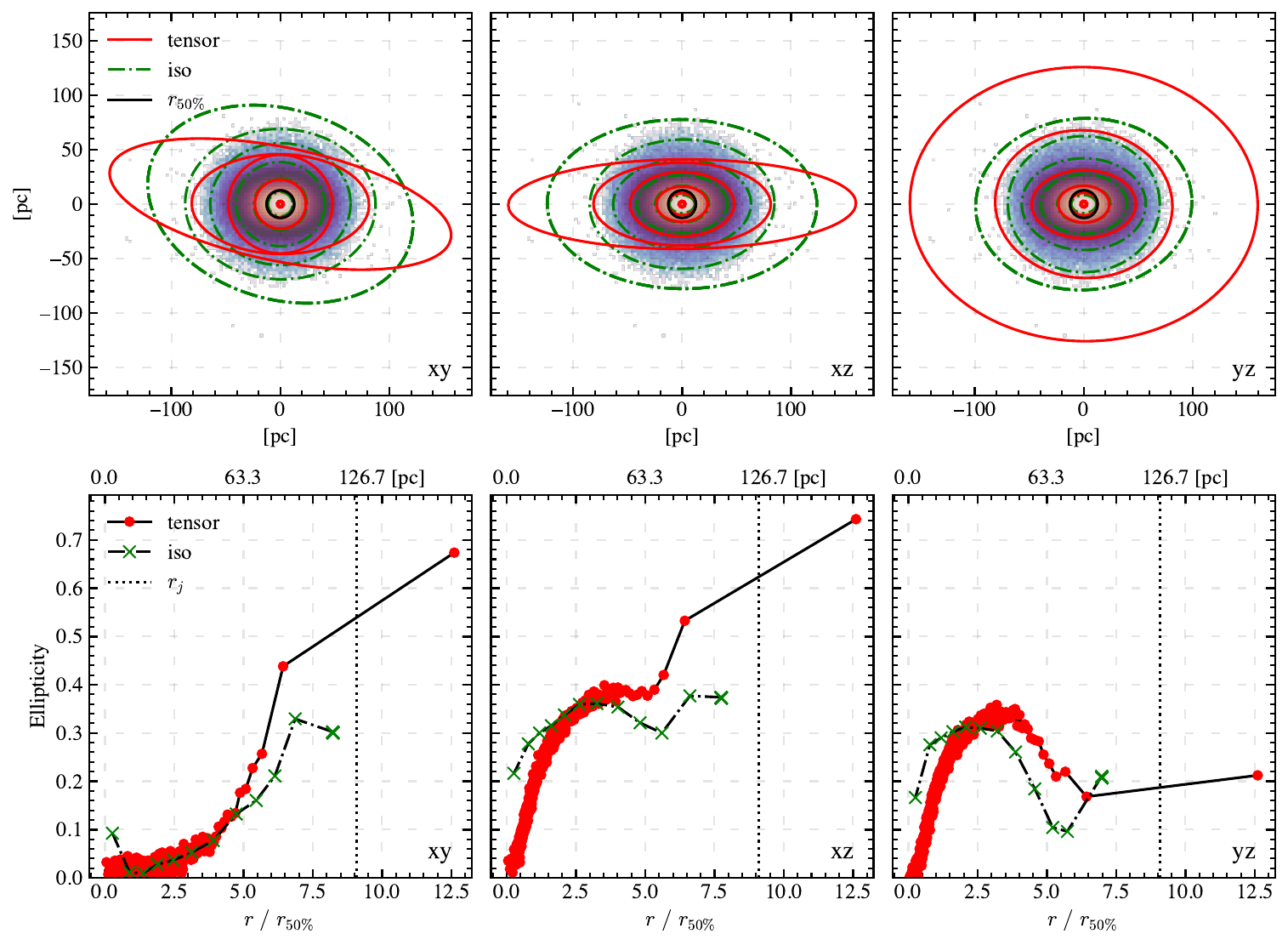}
    \caption{Comparison between the tensor method (red, solid lines) and the iso-density method (green, dash-dotted lines). The first row shows ellipses of the measured ellipticity of respective method overplotted on a 2D histogram of the \texttt{1.5M-A-R4-10} simulation at 12 Gyr, projected in the xy, xz, and yz planes, respectively. The black circle shows where the 3D half-mass radius is. The second row shows the measured ellipticity radial profile (bottom axis in half-mass radii, top-axis in pc) for respective projection. The dotted line indicates the Jacobi radius. We have omitted plotting each ellipse in the first row to avoid cluttering the image.}
    \label{fig: tensor vs iso method}
\end{figure*}

\fix{We compare the morphology of \texttt{1.5M-A-R4-10} as derived using the iso-density contour method and the tensor method in Fig. \ref{fig: tensor vs iso method}. \texttt{1.5M-A-R4-10} is the most massive simulation in the \texttt{ROLLIN'} suite, and is characterised by 1.5M stars. We choose to demonstrate our two methodologies with this simulation, because its morphological measurements should be less affected by low number statistics. The three panels in the top row show the model at 12 Gyr projected onto the three cartesian planes, respectively. On top of the projected distribution we plot the ellipses measured with the iso-density method in green dash-dotted lines, and with the tensor method in red solid lines. The bottom row shows the measured ellipticity at a given distance from the centre. We obtain the radial variation of the flattening with the tensor-method by considering stars inside equally populated spherical shells of 10k stars. Ellipticity measured from the tensor-method (red dots, solid lines) is placed at the average radius of the stars within the corresponding bin, while ellipticity measured with the iso-density method (green cross, dash-dotted line) is placed at the effective radius $\sqrt{ab}$. We use different radius definitions because they are the natural scales for each estimator: the mean radius directly represents the physical extent of the stars used in the tensor calculation, whereas $\sqrt{ab}$ is the unique geometric scale of an iso-density contour. Despite this difference, both values refer to the same radial region of the cluster, ensuring a meaningful comparison of the ellipticity profiles. The iso-density ellipses are obtained by numerically solving eqn. (\ref{eqn: rotated ellipse}) from contours of the projected background distribution(s) shown in the top row.}

\fix{Interestingly, though not surprisingly, the radial profiles for the two methods are not identical. In the $xy$-projection, where the line of sight is aligned with the rotation axis, both methods yield nearly circular central and intermediate regions, and an increasing flattening with radius. Apart from the most central region, the two methods mostly agree up until the Jacobi radius (indicated by the dotted black line). 
In the other two projections, the discrepancy becomes more pronounced: the iso-density method infers enhanced central flattening within the half-mass radius $r_{50\%}$, a maximum flattening around $\sim3 ~ r_{50\%}$, and a local minima around $5-6 ~ r_{50\%}$. In the $xz$ projection, the outer-most region, closer to the Jacobi radius, is found to be the most flattened ($\lesssim 0.4$), whereas the $yz$ projection reports an enhanced, but moderate, flattening ($\lesssim 0.2$). Conversely, the tensor methods measures a non-flattened central region (within $r_{50\%}$), which gradually increases towards a maximum flattening around $4 ~ r_{50\%}$. The overall shape of the radial profile is then mostly the same between the two methods for the two projections: $xz$ shows a significant increase in ellipticity towards the Jacobi radius ($\gtrsim 0.5$), while in the $yz$ plane, there is first a significant decrease ($\lesssim 0.2$), before a moderate increase close to the Jacobi radius. However, the tensor method consistently show a more significant flattening throughout the outer parts, which is consistent with it capturing outer elongations and tidal asymmetries that contribute less to the surface density but dominate the spatial mass distribution. The measured values for the outer-most points (corresponding to the outer-most ellipses in the top row) can be biased by stars near the outer edge of each shell, since the radial bins contain a fixed number of stars and therefore span larger physical distances at larger radii \citep{Freour+2026}.}

\fix{From numerous tests varying the resolution of the projected two-dimensional distributions from which the iso-density contours are defined, as well as the number of stars considered in each bin for the tensor-method, we found that the former is much more sensitive to the specific choice of parameters. For this reason, and the discussion above, we adopt the tensor-method to quantify the morphology of our simulated clusters at different evolutionary stages. This allows us to more easily track temporal changes in their intrinsic axis ratios, triaxiality, and orientation.}

\section{Ellipticity and triaxiality evolution} \label{Appendix: Full Triaxiality}

\begin{figure*}[htbp]
    \centering
    \includegraphics[width=0.98\textwidth]{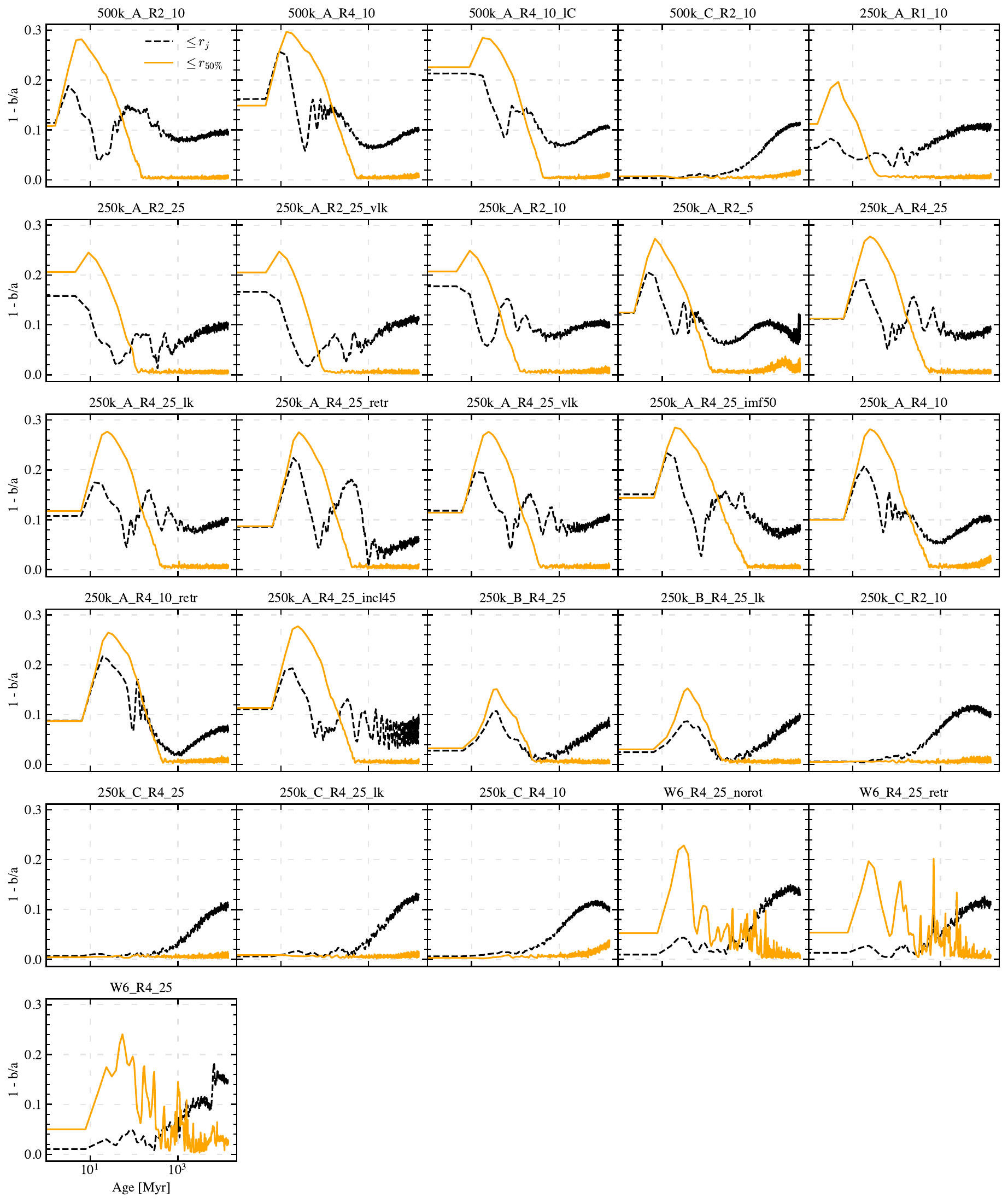}
    \caption{$1-b/a$ as a function of time for all remaining simulations in the \texttt{ROLLIN'} suite.}
    \label{fig: Full 1-b/a}
\end{figure*}

\begin{figure*}[htbp]
    \centering
    \includegraphics[width=0.98\textwidth]{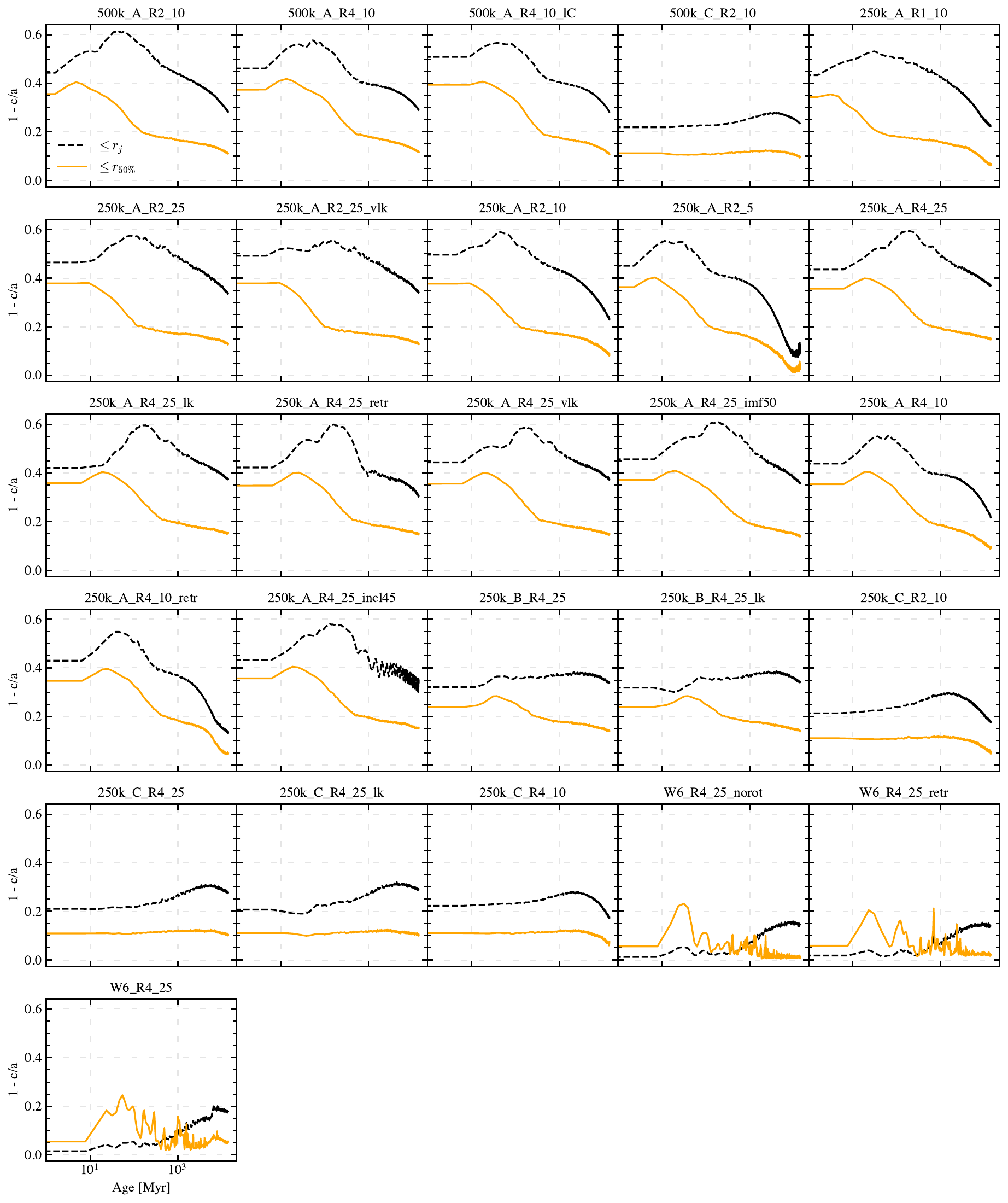}
    \caption{$1-c/a$ as a function of time for all remaining simulations in the \texttt{ROLLIN'} suite.}
    \label{fig: Full 1-c/a}
\end{figure*}

\begin{figure*}[htbp]
    \centering
    \includegraphics[width=0.98\textwidth]{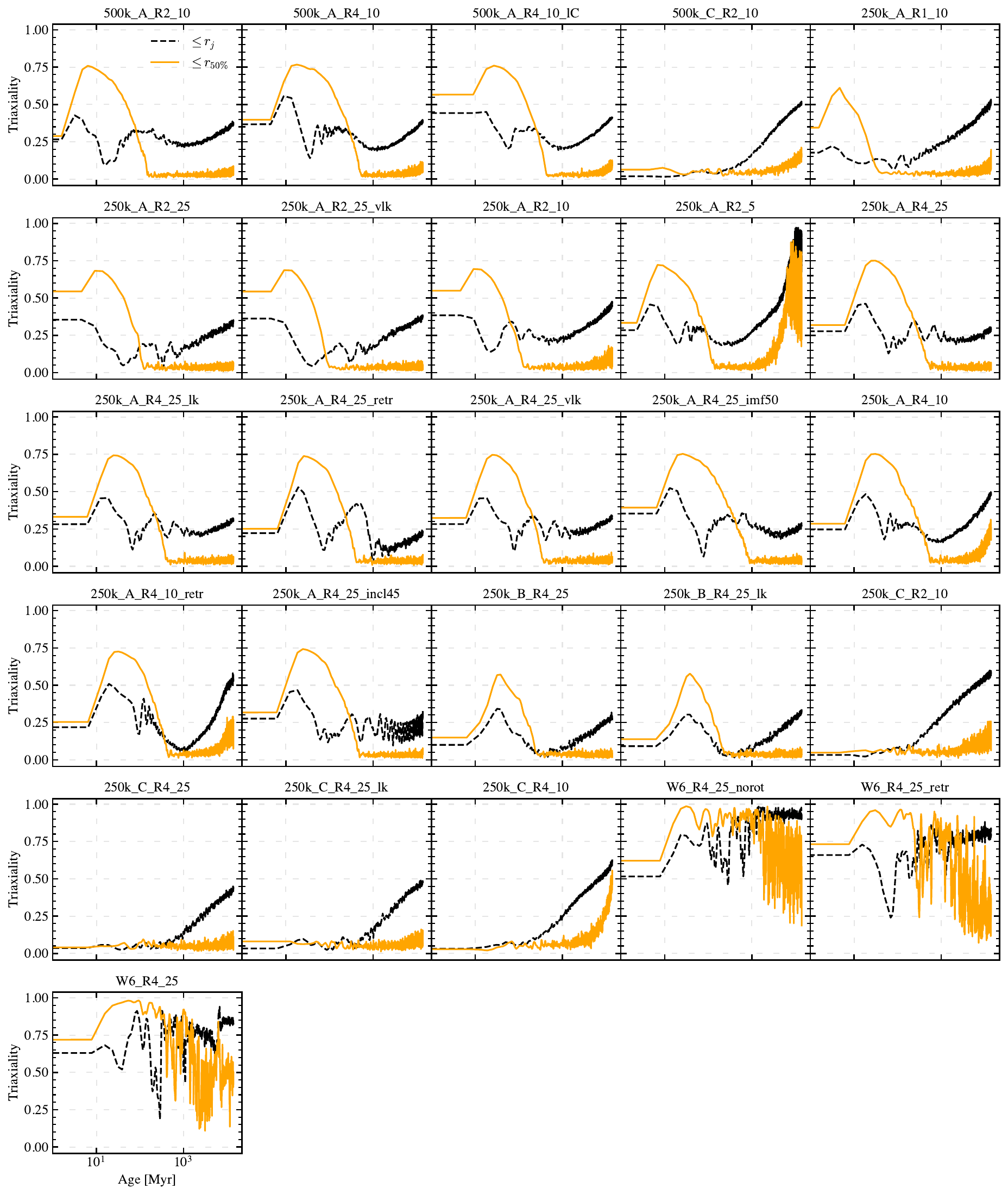}
    \caption{Triaxiality as a function of time for all remaining simulations in the \texttt{ROLLIN'} suite.}
    \label{fig: Full Triaxiality}
\end{figure*}

Temporal evolution of \fix{ellipticity ($1-b/a$ in Fig. \ref{fig: Full 1-b/a} and $1-c/a$ in Fig. \ref{fig: Full 1-c/a})} and triaxiality (Fig. \ref{fig: Full Triaxiality})  within the Jacobi radius (black line) and within the half-mass radius (orange line) for all models in the \texttt{ROLLIN'} suite except for \texttt{1.5M-A-R4-10} which is shown in Fig. \ref{fig: axes ratios evolution}. Table \ref{tab: Morphology} shows each simulations' ellipticity (for the three principal axes) and triaxiality within the Jacobi radius and within the half-mass radius at 12 Gyr. \fix{The last two columns indicate the initial and final (12 Gyr) dynamical timescale within the half-mass radius, calculated using eqn. (\ref{eqn: tdyn}).
}

\begin{table*}[b]
\centering
\caption{Morphological properties of the \texttt{ROLLIN'} simulations at 12 Gyr.}
\begin{adjustbox}{max width=\textwidth}
\label{tab: Morphology}
\begin{tabular}{lcccccccccc}  
\hline \hline
Name & $1-b/a (<r_j)$ & $1-c/a (<r_j)$ & $1-c/b (<r_j)$ & $T (<r_j)$ & $1-b/a (<r_{50\%})$ & $1-c/a (<r_{50\%})$ & $1-c/b (<r_{50\%})$ & $T (<r_{50\%})$ & $t^\mathrm{dyn}_i~[\mathrm{Myr}]$ & $t^\mathrm{dyn}_f~[\mathrm{Myr}]$ \\
\hline
\texttt{1.5M-A-R4-10} & 0.08 & 0.34 & 0.28 & 0.28 & 0.00 & 0.14 & 0.13 & 0.03 & 0.18 & 1.33 \\
\texttt{500k-A-R2-10} & 0.10 & 0.30 & 0.22 & 0.37 & 0.01 & 0.12 & 0.11 & 0.10 & 0.11 & 1.49 \\
\texttt{500k-A-R4-10} & 0.10 & 0.30 & 0.23 & 0.37 & 0.01 & 0.13 & 0.12 & 0.10 & 0.31 & 2.89 \\
\texttt{500k-A-R4-10-lC} & 0.11 & 0.30 & 0.22 & 0.39 & 0.01 & 0.12 & 0.11 & 0.08  & 0.31 & 3.18 \\
\texttt{500k-C-R4-10} & 0.11 & 0.25 & 0.15 & 0.49 & 0.01 & 0.10 & 0.09 & 0.12 & 0.32 & 3.25 \\ %
\texttt{250k-A-R1-10} & 0.10 & 0.23 & 0.14 & 0.48 & 0.01 & 0.07 & 0.06 & 0.10 & 0.06 & 1.12 \\
\texttt{250k-A-R2-25} & 0.10 & 0.35 & 0.28 & 0.32 & 0.00 & 0.13 & 0.13 & 0.03 & 0.16 & 2.25\\
\texttt{250k-A-R2-25-vlk} & 0.11 & 0.35 & 0.27 & 0.36 & 0.00 & 0.13 & 0.13 & 0.03 & 0.16 & 3.32 \\
\texttt{250k-A-R2-10} & 0.10 & 0.25 & 0.17 & 0.43 & 0.01 & 0.10 & 0.09 & 0.10 & 0.16 & 1.64 \\
\texttt{250k-A-R2-5} & 0.07 & 0.08 & 0.01 & 0.86 & 0.00 & 0.02 & 0.02 & 0.27 & 0.16 & 1.64 \\
\texttt{250k-A-R4-25} & 0.09  & 0.37 & 0.32 & 0.27 & 0.01 & 0.15 & 0.15 & 0.05 & 0.44 & 4.30 \\ 
\texttt{250k-A-R4-25-imf50} & 0.09 & 0.37 & 0.31 & 0.28 & 0.00 & 0.14 & 0.14 & 0.03 & 0.45 & 2.84 \\ 
\texttt{250k-A-R4-25-lk} & 0.10 & 0.38 & 0.31 & 0.31 & 0.01 & 0.15 & 0.15 & 0.05 & 0.44 & 5.00 \\
\texttt{250k-A-R4-25-vlk} & 0.11 & 0.39 & 0.31 & 0.32 & 0.01 & 0.15 & 0.15 & 0.04 & 0.44 & 5.42 \\ 
\texttt{250k-A-R4-25-retr} & 0.06 & 0.32 & 0.28 & 0.21 & 0.01 & 0.16 & 0.15 & 0.08 & 0.45 & 4.12 \\
\texttt{250k-A-R4-25-incl45} & 0.05 & 0.32 & 0.28 & 0.20 & 0.01 & 0.15 & 0.15 & 0.04 & 0.44 & 4.24 \\
\texttt{250k-A-R4-10} & 0.10 & 0.24 & 0.16 & 0.46 & 0.02 & 0.10 & 0.08 & 0.17 & 0.44 & 3.96 \\
\texttt{250k-A-R4-10-retr} & 0.07 & 0.15 & 0.08 & 0.49 & 0.01 & 0.05 & 0.04 & 0.16 & 0.45 & 4.13 \\
\texttt{250k-B-R4-25} & 0.08 & 0.35 & 0.29 & 0.27 & 0.00 & 0.14 & 0.14 & 0.01 & 0.44 & 3.94 \\
\texttt{250k-B-R4-25-lk} & 0.09 & 0.35 & 0.29 & 0.29 & 0.01 & 0.15 & 0.14 & 0.07 & 0.44 & 4.53 \\
\texttt{250k-C-R2-10} & 0.10 & 0.19 & 0.10 & 0.56 & 0.00 & 0.06 & 0.05 & 0.06 & 0.16 & 1.95 \\
\texttt{250k-C-R4-25} & 0.11 & 0.29 & 0.20 & 0.43 & 0.00 & 0.11 & 0.10 & 0.04 & 0.44 & 4.95 \\
\texttt{250k-C-R4-25-lk} & 0.13 & 0.30 & 0.20 & 0.47 & 0.01 & 0.10 & 0.09 & 0.10 & 0.44 & 5.80 \\
\texttt{250k-C-R4-10} & 0.11 & 0.20 & 0.10 & 0.58 & 0.03 & 0.07 & 0.05 & 0.40 & 0.44 & 4.84 \\
\texttt{250k-W6-R4-25} & 0.15 & 0.18 & 0.04 & 0.84 & 0.03 & 0.06 & 0.03 & 0.49  & 0.46 & 4.74 \\
\texttt{250k-W6-R4-25-retr} & 0.11 & 0.14 & 0.03 & 0.80 & 0.01 & 0.02 & 0.02 & 0.32 & 0.45 & 5.68 \\
\texttt{250k-W6-R4-25-norot} & 0.14 & 0.15 & 0.02 & 0.92 & 0.01 & 0.01 & 0.00 & 0.92 & 0.46 & 5.70 \\
\hline
\end{tabular}
\end{adjustbox}
\tablefoot{Ellipticity \fix{(along the three principal axes)} and triaxiality within the Jacobi radius \fix{$r_j$} and within the half-mass \fix{$r_{50\%}$} radius at 12 Gyr for all \texttt{ROLLIN'} models. \fix{The last two columns show the initial and final dynamical timescale (in Myr) of the models within the half-mass radius.}}
\end{table*}

\section{Bar properties} \label{Appendix: Bar properties}

Here we include the full comparison for the properties considered in Fig. \ref{fig: bar vs ICs}. Panels (b) and (c) further show that neither the dynamical timescale, nor relaxation timescale, plays a role for the peak strength. Panels (d) and (f) show that the rotational support and relaxation time do not affect the time of formation, given that a bar forms, and panels (g) and (h) show that the lifetime of the bar is not dependent on the rotational support or the dynamical timescale.

\begin{figure*}[htbp!]
    \centering
    \sidecaption
    \includegraphics[width=12cm]{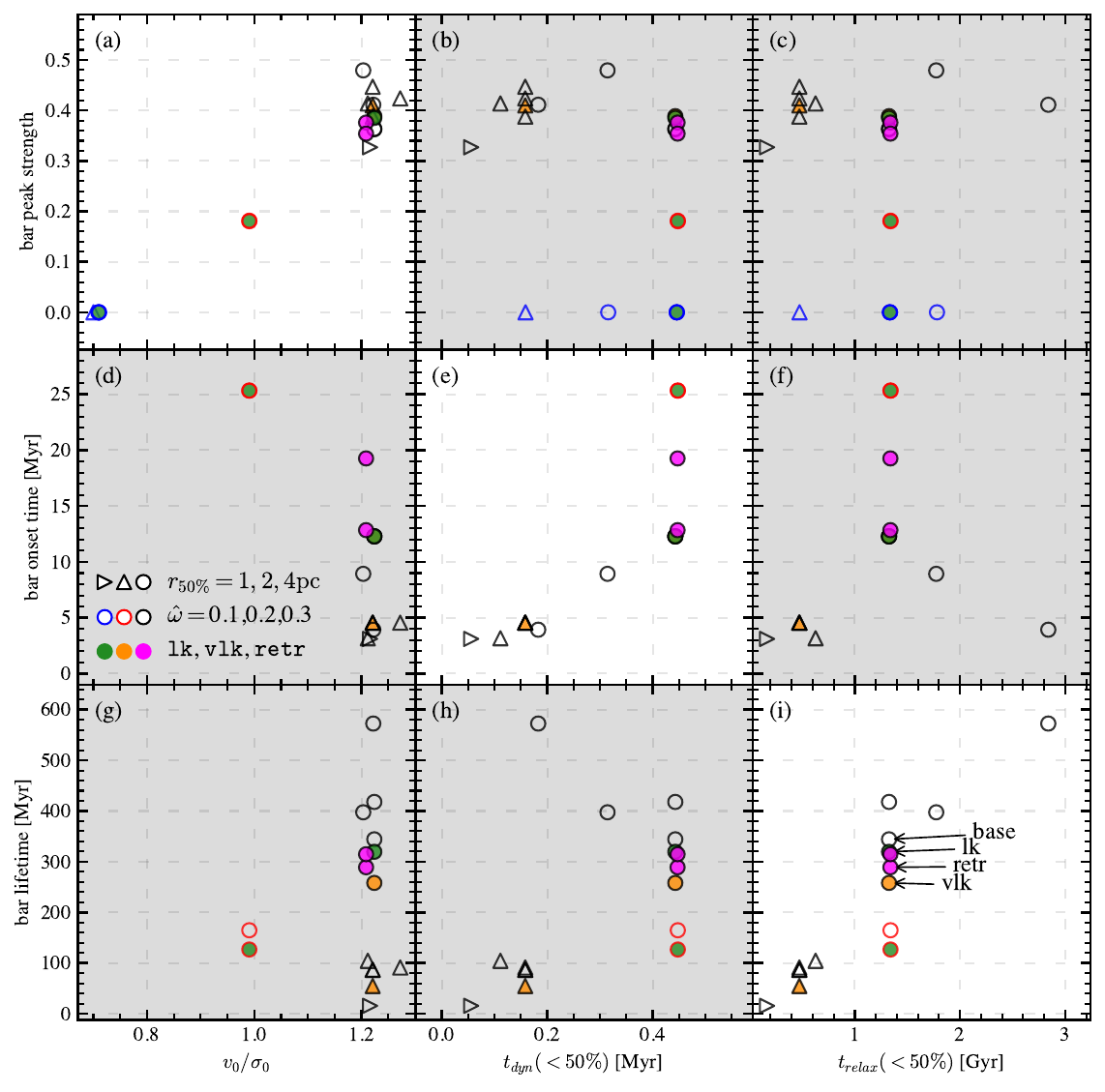}
    \caption{Bar properties (peak strength, onset time, and lifetime) for the models and how they depend on a subset of initial conditions: rotational support $v/\sigma$ (first column), dynamical timescale in Myr (second column), and relaxation timescale (third column). Panels with white backgrounds highlight the correlations between bar properties and initial conditions (see Fig. \ref{fig: bar vs ICs}), while grey backgrounds indicate no correlation. Different symbols indicate models with different initial half-mass radii, different edge-colours indicate the initial rotation strength, and the different filling colours indicate models with any of \texttt{-lk}, \texttt{-retr}, or \texttt{-vlk}.}
    \label{fig: bar vs ICs full}
\end{figure*}
\newpage
\section{Bars without stellar evolution} \label{Appendix: Bars}

\begin{figure}[htbp!]
    \centering
    \includegraphics[width=0.95\textwidth]{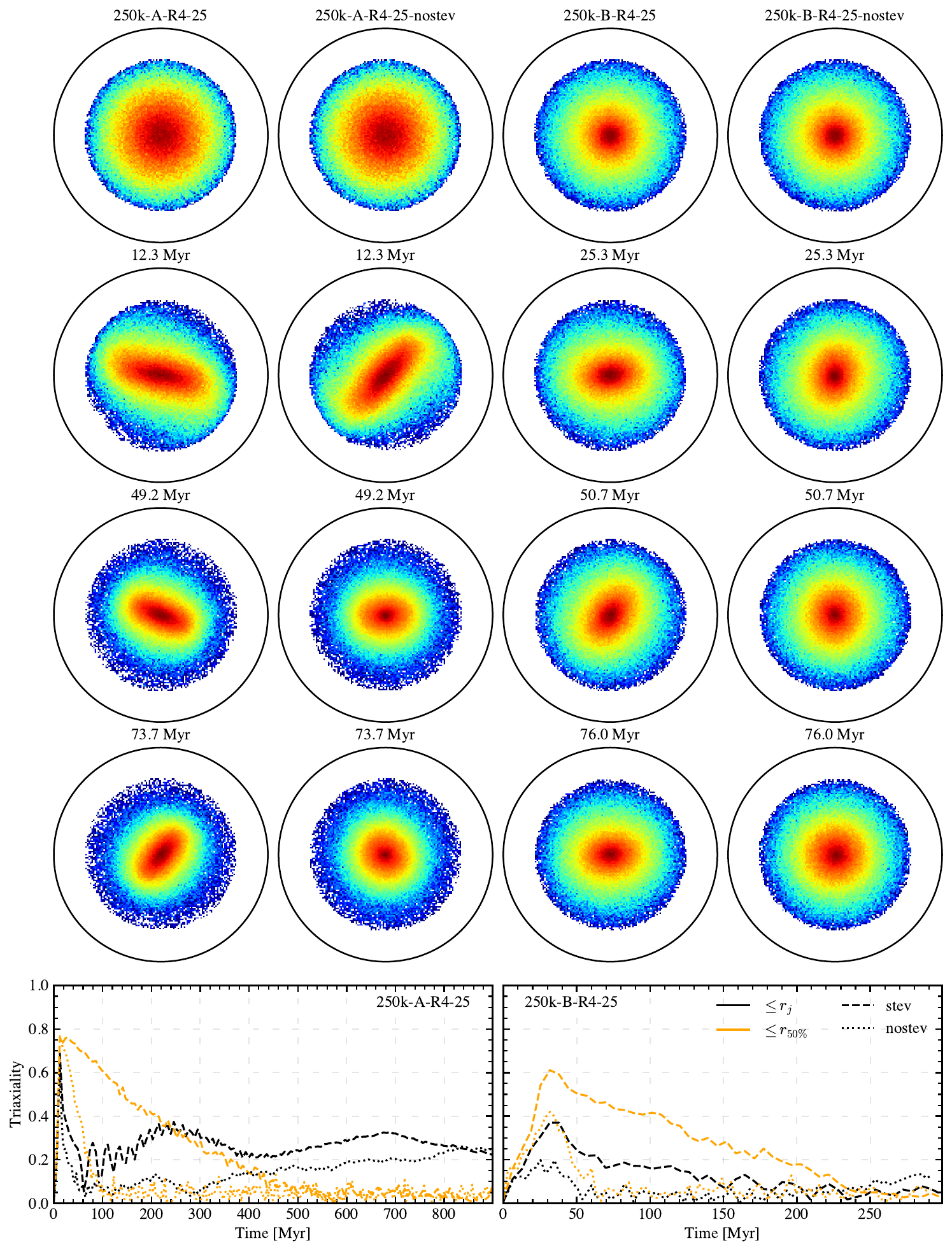}
    \caption{Snapshots of \texttt{250k-A-R4-25}, \texttt{250k-A-R4-25-nostev}, \texttt{250k-B-R4-25}, and \texttt{250k-B-R4-25-nostev} at different times, and their overlapping evolution for triaxiality.}
    \label{fig: StevNoStev}
\end{figure}

Fig. \ref{fig: StevNoStev} shows the two control models \texttt{250k-A-R4-25} and \texttt{250k-B-R4-25} which we ran without stellar evolution (\texttt{250k-A-R4-25-nostev} and \texttt{250k-B-R4-25-nostev}, respectively). First row shows the initial distribution at $t=0$ viewed along the rotation axis, with subsequent rows showing the projection at different times (indicated in the figure). A bar-like structure forms for each simulation, but they are quickly destroyed without stellar evolution. The last row shows the temporal evolution of the triaxiality for each model.

\section{Remaining correlation panels} \label{Appendix: Remaining Correlations}

\begin{figure*}[htbp!]
    \sidecaption
    \includegraphics[width=12cm]{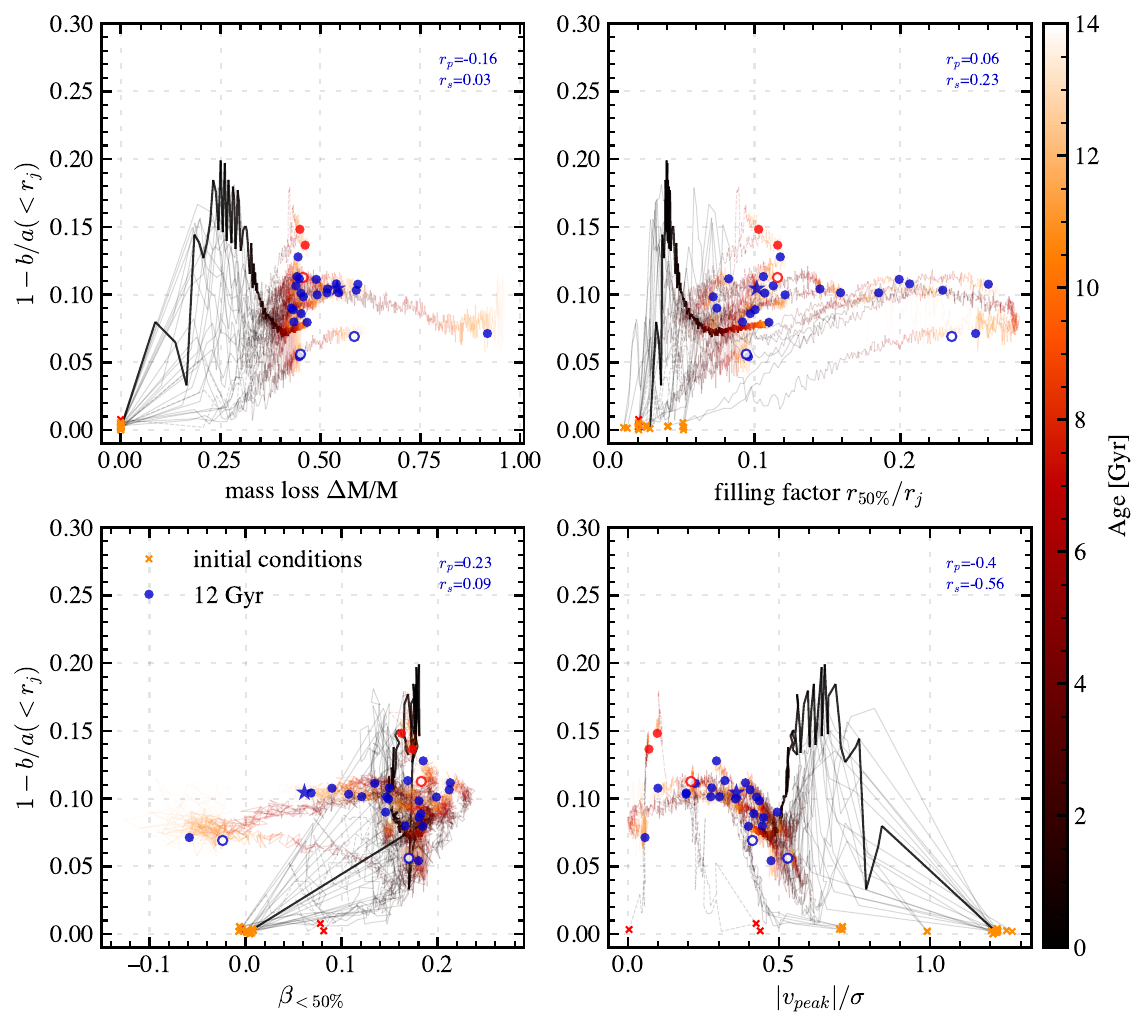}
    \caption{Same as Fig. \ref{fig: exz vs stuff} but for the $1-b/a$ projection. The Pearson and Spearman coefficients when including the Wilson models are for mass loss $r_p=-0.22$, $r_s=-0.03$; filling factor $r_p=-0.04$, $r_s=0.19$; anisotropy parameter $r_p=0.26$, $r_s=0.14$; and rotational support $r_p=-0.59$, $r_s=-0.63$.}
    \label{fig: exy vs stuff}
\end{figure*}

\begin{figure*}[htbp]
    \sidecaption
    \includegraphics[width=12cm]{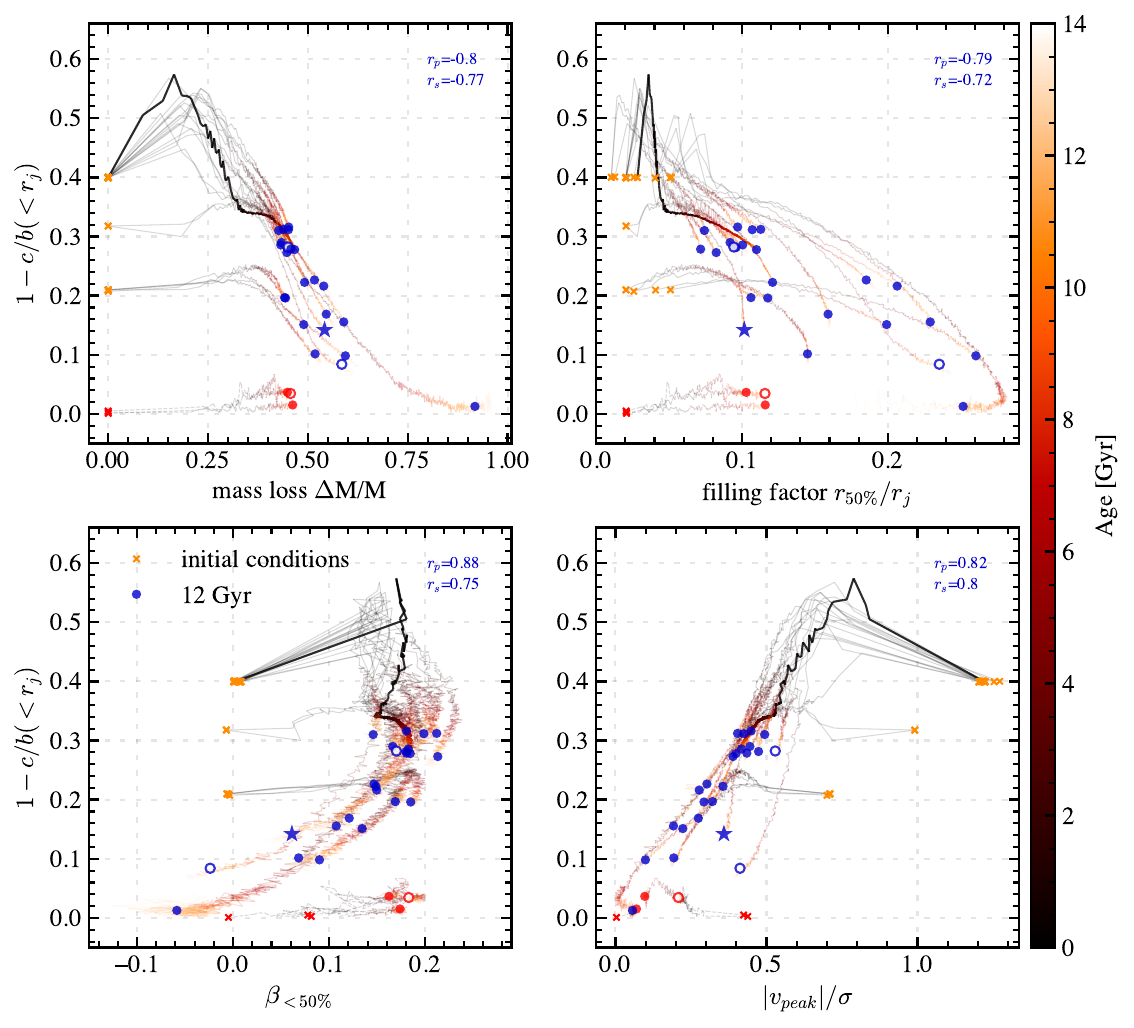}
    \caption{Same as Fig. \ref{fig: exz vs stuff} but for the $1-c/b$ projection. The Pearson and Spearman coefficients when including the Wilson models are for mass loss $r_p=-0.54$, $r_s=-0.63$; filling factor $r_p=-0.53$, $r_s=-0.63$; anisotropy parameter $r_p=0.60$, $r_s=0.55$; and rotational support $r_p=0.87$, $r_s=0.84$.}
    \label{fig: eyz vs stuff}
\end{figure*}

\begin{figure*}[htbp!]
    \sidecaption 
    \includegraphics[width=12cm]{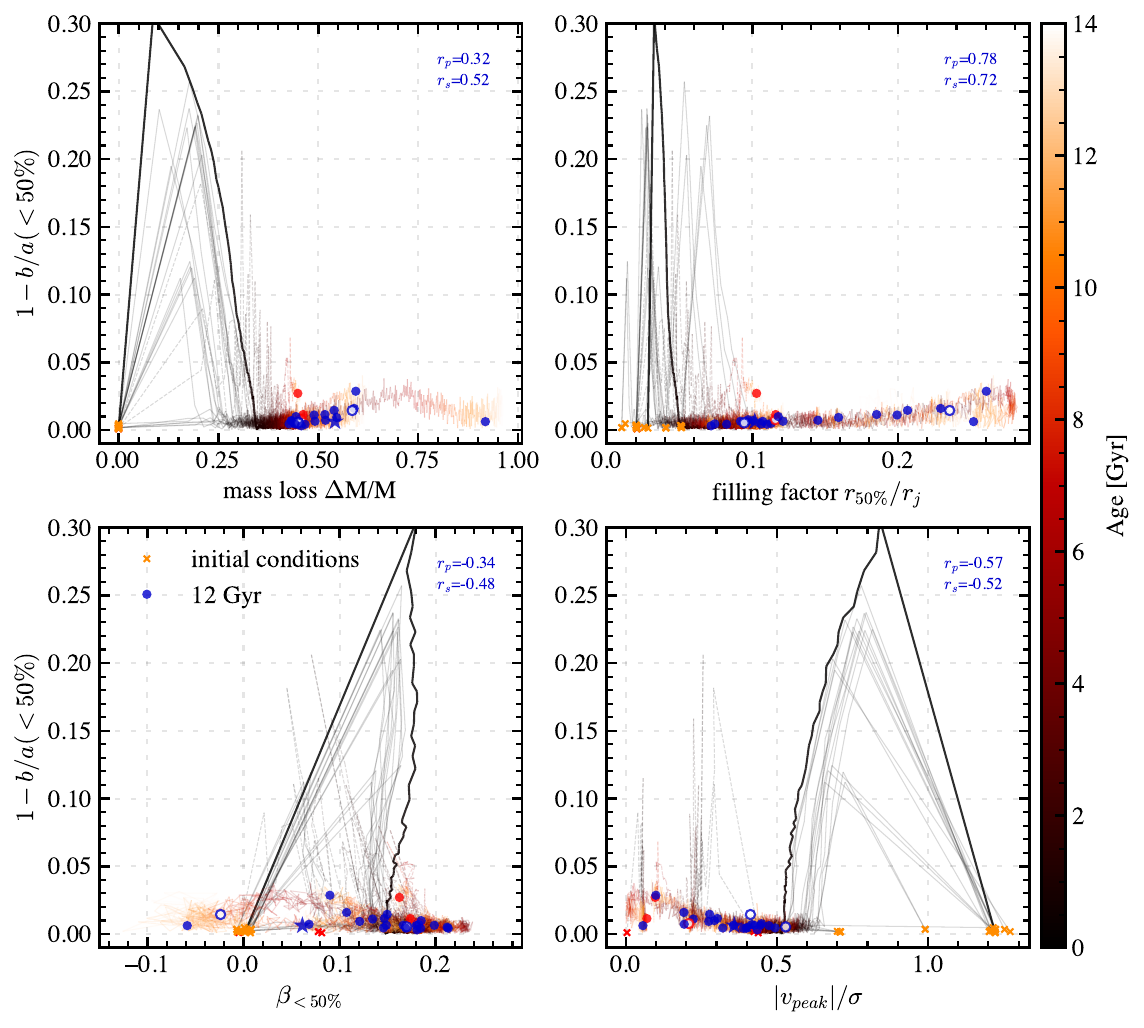}
    \caption{Same as Fig. \ref{fig: exy vs stuff} but within the half-mass radius. The Pearson and Spearman coefficients when including the Wilson models are for mass loss $r_p=0.16$, $r_s=0.41$; filling factor $r_p=0.54$, $r_s=0.62$; anisotropy parameter $r_p=-0.17$, $r_s=-0.40$; and rotational support $r_p=-0.66$, $r_s=-0.62$.}
    \label{fig: exy50 vs stuff}
\end{figure*}

\begin{figure*}[htbp!]
    \sidecaption
    \includegraphics[width=12cm]{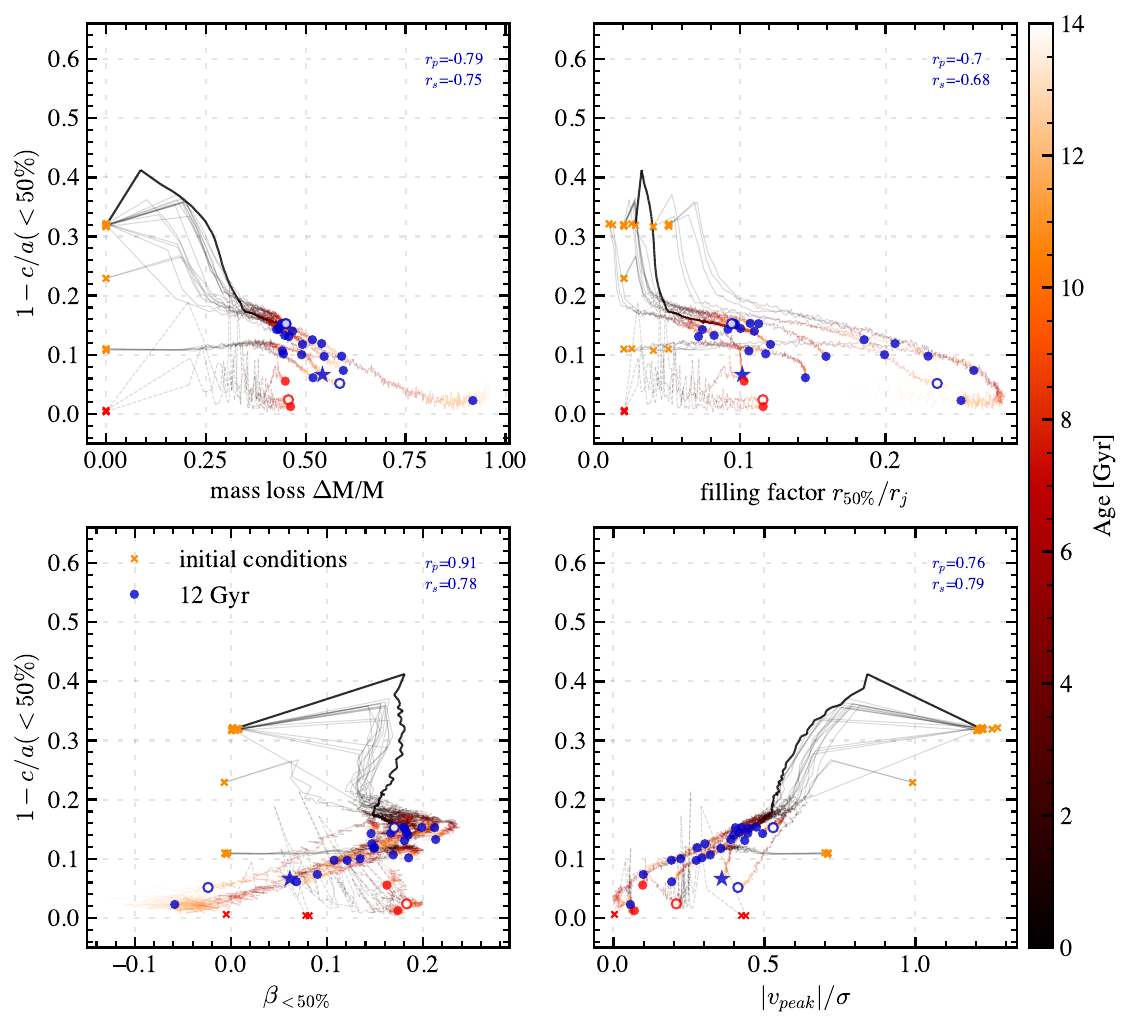}
    \caption{Same as Fig. \ref{fig: exz vs stuff} but within the half-mass radius. The Pearson and Spearman coefficients when including the Wilson models are for mass loss $r_p=-0.57$, $r_s=-0.63$; filling factor $r_p=-0.45$, $r_s=-0.55$; anisotropy parameter $r_p=0.67$, $r_s=0.67$; and rotational support $r_p=0.90$, $r_s=0.89$.}
    \label{fig: exz50 vs stuff}
\end{figure*}

\begin{figure*}[htbp!]
    \sidecaption
    \includegraphics[width=12cm]{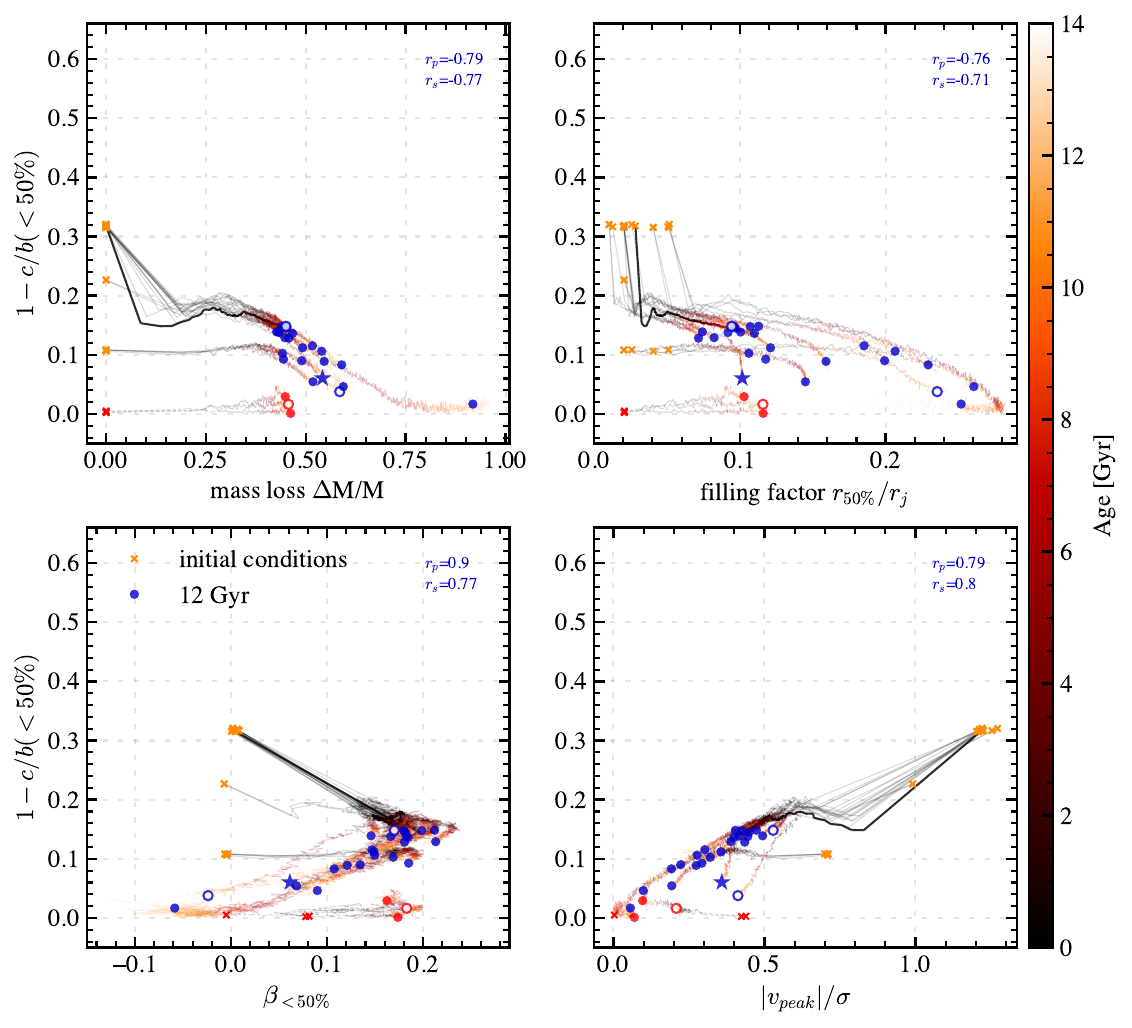}
    \caption{Same as Fig. \ref{fig: eyz vs stuff} but within the half-mass radius. The Pearson and Spearman coefficients when including the Wilson models are for mass loss $r_p=-0.56$, $r_s=-0.65$; filling factor $r_p=-0.49$, $r_s=-0.58$; anisotropy parameter $r_p=0.64$, $r_s=0.66$; and rotational support $r_p=0.93$, $r_s=0.91$.}
    \label{fig: eyz50 vs stuff}
\end{figure*}

\begin{figure*}[htbp!]
    \sidecaption
    \includegraphics[width=12cm]{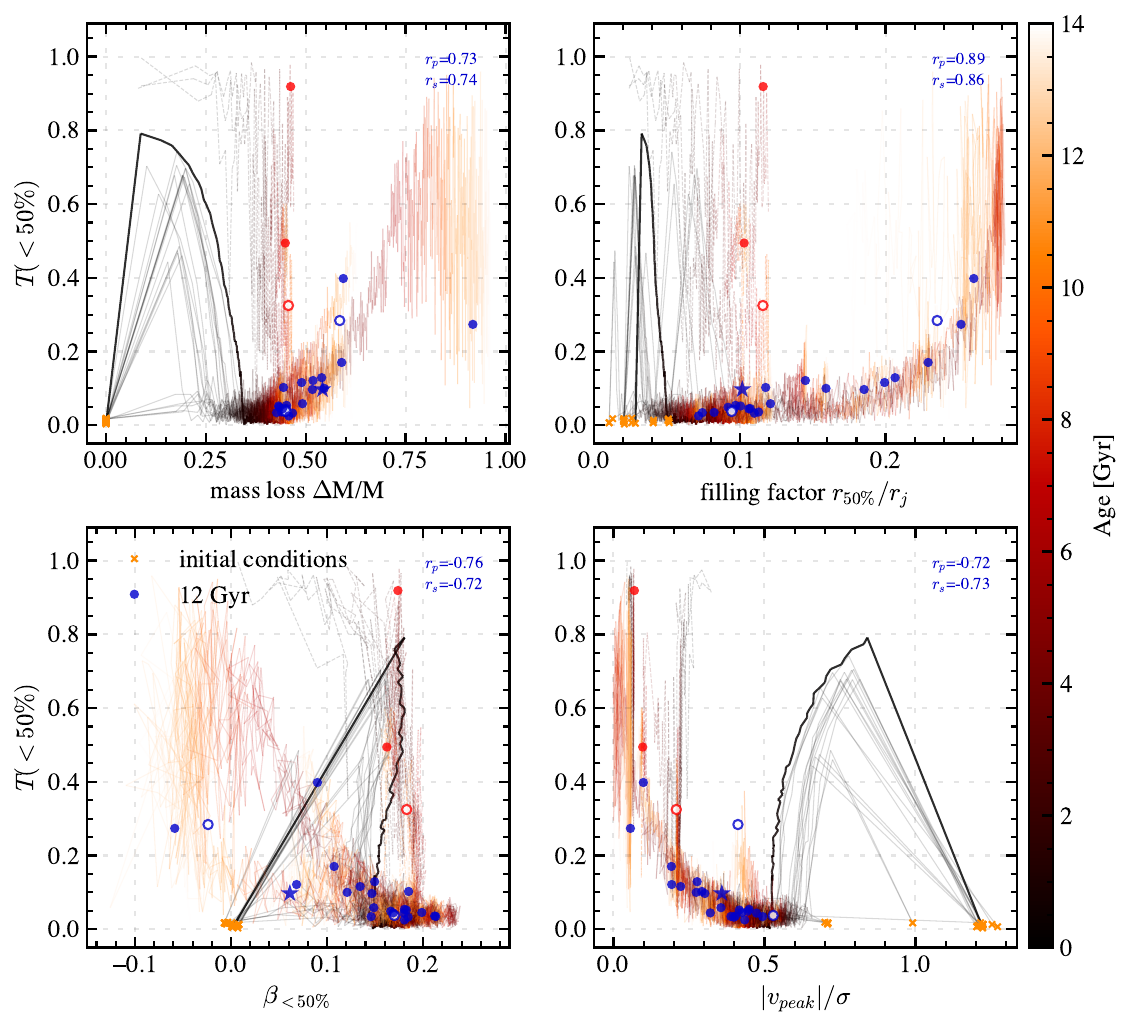}
    \caption{Same as Fig. \ref{fig: T vs stuff} but within the half-mass radius. The Pearson and Spearman coefficients when including the Wilson models are for mass loss $r_p=0.19$, $r_s=0.59$; filling factor $r_p=0.25$, $r_s=0.71$; anisotropy parameter $r_p=-0.19$, $r_s=-0.59$; and rotational support $r_p=-0.76$, $r_s=-0.87$.}
    \label{fig: T50 vs stuff}
\end{figure*}

Correlations between flattening along the principal axes within the Jacobi radius (Figs. \ref{fig: exy vs stuff} \& \ref{fig: eyz vs stuff}), within the half mass radius (Figs. \ref{fig: exy50 vs stuff}-\ref{fig: eyz50 vs stuff}), and for triaxiality within the half-mass radius (Fig. \ref{fig: T50 vs stuff}).

\end{appendix}
\end{document}